\documentclass[]{interact}
\usepackage{lscape}
\usepackage{multirow}
\usepackage{framed}
\usepackage{longtable}
\usepackage{makecell}
\usepackage{epstopdf}
\usepackage[caption=false]{subfig}

\usepackage[natbibapa,nodoi]{apacite}
\theoremstyle{plain}

\theoremstyle{definition}

\theoremstyle{remark}

\begin{document}


\title{How should my chatbot interact? A survey on social characteristics in human-chatbot interaction design (preprint)}

\author{
\name{Ana Paula Chaves\textsuperscript{a,b}\thanks{CONTACT Ana Paula Chaves. Email: anachaves@utfpr.edu.br} and Marco Aurelio Gerosa\textsuperscript{a}}
\affil{\textsuperscript{a}School of Informatics, Computing, and Cyber Systems, Northern Arizona University, Flagstaff, Arizona, USA; \textsuperscript{b}Federal University of Technology--Paran\'a, Campo Mour\~ao, Paran\'a, Brazil}
}

\maketitle

\begin{abstract}
Chatbots' growing popularity has brought new challenges to HCI, having changed the patterns of human interactions with computers. The increasing need to approximate conversational interaction styles raises expectations for chatbots to present social behaviors that are habitual in human-human communication. In this survey, we argue that chatbots should be enriched with social characteristics that cohere with users' expectations, ultimately avoiding frustration and dissatisfaction. We bring together the literature on disembodied, text-based chatbots to derive a conceptual model of social characteristics for chatbots. We analyzed 56 papers from various domains to understand how social characteristics can benefit human-chatbot interactions and identify the challenges and strategies to designing them. Additionally, we discussed how characteristics may influence one another. Our results provide relevant opportunities to both researchers and designers to advance human-chatbot interactions.
\end{abstract}

\begin{keywords}
chatbots; social characteristics; human-chatbot interaction; survey
\end{keywords}

\section{Introduction}
\label{sec:introduction}

Chatbots are computer programs that interact with users in natural language \citep{shawar2007chatbots}. The origin of the chatbot concept dates back to 1950 \citep{turing1950computing}. ELIZA \citep{weizenbaum1966eliza} and A.L.I.C.E. \citep{wallace2009anatomy} are examples of early chatbot technologies, where the main goal was to mimic human conversations. Over the years, the chatbot concept has evolved. Today, chatbots can have distinct and diverse characteristics, which has resulted in several synonyms, such as multimodal agents, chatterbots, and conversational interfaces. In this survey, we use the term ``\textit{chatbot}'' to refer to \textit{a disembodied conversational agent that holds a natural language conversation via text-based environment to engage the user in either a general-purpose or task-oriented conversation.}

Chatbots are changing the patterns of interactions between humans and computers \citep{folstad2017chatbots}. Many instant messenger tools, such as Skype, Facebook Messenger, and Telegram provide platforms to develop and deploy chatbots, which either engage with users in general conversations or help them solve domain specific tasks \citep{dale2016return}. As messaging tools increasingly become platforms, traditional websites and apps are providing space for this new form of human-computer interaction (HCI) \citep{folstad2017chatbots}. For example, in the 2018 F8 Conference, Facebook announced that it had 300K active chatbots on Facebook Messenger \citep{boiteux2018messenger}. The BotList \footnote{https://botlist.co/} website indexes thousands of chatbots for education, entertainment, games, health, productivity, travel, fun, and several other categories. The growth of chatbot technology is changing how companies engage with their customers \citep{gnewuch2017towards, brandtzaeg2018chatbots}, students engage with their learning groups \citep{hayashi2015social, tegos2016investigation}, and patients self-monitor the progress of their treatment \citep{fitzpatrick2017delivering}, among many other applications.

However, chatbots still fail to meet users' expectations \citep{luger2016like, jain2018evaluating, brandtzaeg2018chatbots, zamora2017sorry}. While many studies on chatbot design focus on improving chatbots' functional performance and accuracy (see, e.g., \citet{jiang2017towards, maslowski2017wild}), the literature has consistently suggested that chatbots' interactional goals should also include social capabilities \citep{jain2018evaluating, liao2018all}. According to the Media Equation theory \citep{reeves1996people}, people naturally respond to social situations when interacting with computers \citep{nass1994computers, fogg2003computers}. As chatbots are designed to interact with users in a way that mimics person-to-person conversations, new challenges in HCI arise \citep{nguyen2018understanding, folstad2017chatbots}. Neururer and colleagues (\citeyear{neururer2018perceptions}) state that making a conversational agent acceptable to users is primarily a social, not only technical, problem. Studies on chatbots have shown that people prefer agents who: conform to gender stereotypes associated with tasks \citep{forlizzi2007interface}; self-disclose and show reciprocity when recommending \citep{lee2017enhancing}; and demonstrate a positive attitude and mood \citep{thies2017how}. When chatbots do not meet these expectations, the user may experience frustration and dissatisfaction \citep{luger2016like, zamora2017sorry}. In contrast, designing overly humanized agents results in uncanny feelings and increased expectations \citep{gnewuch2017towards, ciechanowski2018shades}, which also negatively impacts the interaction. The challenge remains as to what social characteristics are relevant for improving chatbots' communication and social skills and in which domains they have shown to be beneficial.

Although chatbots' social characteristics have been explored in the literature, this knowledge is spread across several domains in which chatbots have been studied, such as customer services, education, finances, and travel. In the HCI domain, some studies focus on investigating the social aspects of human-chatbot interactions (see, e.g., \citet{ciechanowski2018shades, ho2018psychological, lee2017enhancing}). However, most studies focus on a single or small set of characteristics (e.g., \citet{mairesse2009can, schlesinger2018let}); in other studies, the social characteristics emerged as secondary, exploratory results (e.g., \citet{tallyn2018ethnobot, toxtli2018understanding}). It has become difficult to find evidence regarding what characteristics are important for designing a particular chatbot, and what research opportunities exist in the field.

Whilst the literature has extensively reviewed the technical aspects of chatbots design (e.g., \citet{ramesh2017survey, thorne2017chatbots, winkler2018unleashing, ahmad2018review, walgama2017chatbots, deshpande2017survey, masche2017review}), a lack of studies brings together the social characteristics that influence the way users perceive and behave toward chatbots. To fill this gap, this survey compiles research initiatives for understanding the impact of chatbots' social characteristics on the interaction. We bring together literature that is spread across several research areas. From our analysis of 56 scientific studies, we derive a conceptual model of social characteristics, aiming to help researchers and designers identify the subset of characteristics that are relevant to their context and how adopting--or neglecting--a particular characteristic may influence the way humans perceive the chatbots. The research question that guided our investigation was: \textbf{What chatbot social characteristics benefit human interaction and what are the challenges and strategies associated with them?}

To answer this question, we discuss why designing a chatbot with a particular characteristic can enrich the human-chatbot interaction. Our results provide insight into whether the characteristic is desirable for a particular chatbot, so that designers can make informed decisions by selecting the appropriate subset of characteristics, as well as inspire researchers' further investigations. In addition, we discuss the influence of humanness and the conversational context on users' perceptions as well as the interrelationship among the identified characteristics. We stated 22 propositions about how social characteristics may influence one another. In the next section, we present an overview of the studies included in this survey.

\section{Overview of the surveyed literature}
\label{sec:overview}

The literature presents no coherent definition of chatbots; thus, to find relevant studies we used a search string that includes the synonyms \textit{``chatbots, chatterbots, conversational agents, conversational interfaces, conversational systems, conversation systems, dialogue systems, digital assistants, intelligent assistants, conversational user interfaces, and conversational UI''}. We explicitly left out studies that relate to embodiment (\textit{``ECA, multimodal, robots, eye-gaze, gesture''}), and speech input mode (\textit{``speech-based, speech-recognition, voice-based''}). The search string did not include the term ``social bots'', because it refers to chatbots that produce content for social networks such as Twitter \citep{ferrara2016rise}. We did not include ``personal assistants'' either, since this term consistently refers to commercially available, voice-based assistants such as Google Assistant, Amazon's Alexa, Apple's Siri, and Microsoft's Cortana.

We decided to not include terms that relate to social characteristics/traits, because most studies do not explicitly label their results as such. Additionally, to find studies from a variety of domains (not only computing), we used Google Scholar to search for papers. We started with a set of about one thousand papers. In the first round, we reduced this amount by about half based on the titles. We also removed papers for which full text was not available or written in a language other than English. We started the second round with 464 studies. In this round, we excluded short/position papers, theses, book chapters, and documents published in non-scientific venues. We also read the abstracts and removed all the papers that focused on technical aspects, rather than social ones. The third and last round started with 104 papers. After reading the full texts, we removed the papers that either did not highlight social characteristics or represented previous, ongoing research of another complete study from the same authors. The datasets with all analyzed studies (original list of results and exclusions per round) is available in \citet{chaves2019dataset}.

After filtering the search results, we had 56 remaining studies. Most of the selected studies are recent publications (less than 10 years old). The publication venues include the domains of human-computer interactions (25 papers), learning and education (8 papers), information and interactive systems (8 papers), virtual agents (5 papers), artificial intelligence (3 papers), and natural language processing (2 papers). We also found papers from health, literature \& culture, computer systems, communication, and humanities (1 paper each). Most papers (59\%) focus on task-oriented chatbots. General purpose chatbots reflect 33\% of the surveyed studies. Most general purpose chatbots (16 out of 19) are designed to handle topic-unrestricted conversations. The most representative specific domain is education, with 9 papers, followed by customer services, with 5 papers. See the supplementary materials for the complete list of topics. 

Most surveyed studies adopted real chatbots (35 out of 56); among them, 18 studies analyze logs of conversations or users' perceptions of third-party chatbots such as Cleverbot (e.g., \citet{corti2016co}), Talkbot (e.g., \citet{brahnam2012gender}), and Woebot \citet{fitzpatrick2017delivering}. Nine studies introduce a self-developed architecture and/or dialogue management (e.g. \citet{dohsaka2014effects, kumar2010socially}). In another nine studies, the chatbots were designed for research purposes using third-party platforms for chatbot development and deployment, such as IBM's Watson service (e.g., \citet{liao2018all}) and Microsoft's Bot Framework \citep{toxtli2018understanding} as well as pattern-matching packages such as Artificial Intelligence Markup Language (AIML) \citep{silvervarg2013iterative}. When a chatbot was simulated (11 studies), Wizard of Oz (WoZ) is the most used technique (9 studies). In a WoZ study, participants believe to be interacting with a chatbot when, in fact, a person (or ``wizard'') pretends to be the automated system \citep{dahlback1993wizard}. Eight studies do not address a particular chatbot. See the supplementary materials for details.

We analyzed the papers by searching for chatbot behavior or attributed characteristics that influence the way users perceive it and behave toward it. Noticeably, the characteristics and categories are seldom explicitly highlighted to in the literature, so the conceptual model was derived using a qualitative coding process inspired by methods such as Grounded Theory \citep{auerbach2003qualitative} (open coding stage). For each study (\textit{document}), we selected relevant statements from the paper (\textit{quotes}) and labeled them as a characteristic (\textit{code}). These steps were performed by one researcher (the first author). After coding all the studies, a second researcher (the second author) reviewed the produced set of characteristics and both researchers participated in discussion sessions to identify characteristics that could be merged, renamed, or removed. At the end, the characteristics were grouped into the categories, depending on whether the characteristic relates to the chatbot's virtual representation, conversational behavior, or social protocols. Finally, the quotes for each characteristic were labeled as references to benefits, challenges, or strategies.

We derived a total of 11 social characteristics, and grouped them into three categories, as depicted in Table \ref{tab:conceptualmodel}: \textbf{conversational intelligence}, \textbf{social intelligence}, and \textbf{personification}. The next section describes the derived conceptual model.

\begin{landscape}
\begin{table}[ht]
\scriptsize
\centering
\caption{Conceptual model of chatbots social characteristics. Eleven characteristics found in the literature were grouped into three main categories, depending on whether the social characteristic relates to the chatbot's conversational behavior, social protocols, or virtual representation. For each characteristic, we point out the benefits, challenges, and strategies of implementing the characteristic in chatbots as reported in the literature.}
\vspace{1mm}
\begin{tabular}{c|p{2.5cm}|p{6cm}|p{6cm}|p{6cm}|}
\cline{2-5}
 & \textbf{\shortstack{\\Social\\Characteristics}}& \multicolumn{1}{c|}{\textbf{Benefits}} & \multicolumn{1}{c|}{\textbf{Challenges}} & \multicolumn{1}{c|}{\textbf{Strategies}} \\ \hline
\multicolumn{1}{|c|}{\multirow{14}{*}{\rotatebox{90}{\textbf{\shortstack{Conversational\\Intelligence}}}}} & \multirow{6}{*}{\textit{Proactivity}} & \textbf{[B1]} to provide additional information & \textbf{[C1]} timing and relevance & \textbf{[S1]} to leverage conversational context \\
\multicolumn{1}{|l|}{} &  & \textbf{[B2]} to inspire users and to keep the conversation alive & \textbf{[C2]} privacy & \textbf{[S2]} to select a topic randomly \\
\multicolumn{1}{|l|}{} &  & \textbf{[B3]} to recover from a failure & \textbf{[C3]} users' perception of being controlled &  \\
\multicolumn{1}{|l|}{} &  & \textbf{[B4]} to improve conversation productivity &  &  \\
\multicolumn{1}{|l|}{} &  & \textbf{[B5]} to guide and engage users &  &  \\ \cline{2-5}
\multicolumn{1}{|l|}{} & \multirow{4}{*}{\textit{Conscientiousness}} & \textbf{[B1]} to keep the conversation on track & \textbf{[C1]} to handle task complexity & \textbf{[S1]} conversational flow \\
\multicolumn{1}{|l|}{} &  & \textbf{[B2]} to demonstrate understanding & \textbf{[C2]} to harden the conversation & \textbf{[S2]} visual elements \\
\multicolumn{1}{|l|}{} &  & \textbf{[B3]} to hold a continuous conversation & \textbf{[C3]} to keep the user aware of the chatbot's context & \textbf{[S3]} confirmation messages \\ \cline{2-5}
\multicolumn{1}{|l|}{} & \multirow{4}{*}{\textit{Communicability}} & \textbf{[B1]} to unveil functionalities & \textbf{[C1]} to provide business integration & \textbf{[S1]} to clarify the purpose of the chatbot \\ 
\multicolumn{1}{|l|}{} &  & \textbf{[B2]} to manage the users' expectations & \textbf{[C2]} to keep visual elements consistent with textual inputs & \textbf{[S2]} to advertise the functionality and suggest the next step \\
\multicolumn{1}{|l|}{} &  &  &  & \textbf{[S3]} to provide a help functionality \\\hline
\multicolumn{1}{|l|}{\multirow{19}{*}{\rotatebox{90}{\textbf{\shortstack{Social\\Intelligence}}}} }& \multirow{6}{*}{\textit{Damage control}} & \textbf{[B1]} to appropriately respond to harassment & \textbf{[C1]} to deal with unfriendly users & \textbf{[S1]} emotional reactions \\
\multicolumn{1}{|l|}{} &  & \textbf{[B2]} to deal with testing & \textbf{[C2]} to identify abusive utterances & \textbf{[S2]} authoritative reactions \\
\multicolumn{1}{|l|}{} &  & \textbf{[B3]} to deal with lack of knowledge & \textbf{[C3]} to fit the response to the context & \textbf{[S3]} to ignore the user's utterance and change the topic \\
\multicolumn{1}{|l|}{} &  &  &  & \textbf{[S4]} \textit{conscientiousness} and \textit{communicability} \\
\multicolumn{1}{|l|}{} &  &  &  & \textbf{[S5]} to predict users' satisfaction \\\cline{2-5}
\multicolumn{1}{|l|}{} & \multirow{2}{*}{\textit{Thoroughness}} & \textbf{[B1]} to increase human-likeness & \textbf{[C1]} to decide how much to say & Not identified \\
\multicolumn{1}{|l|}{} &  & \textbf{[B2]} to increase believability & \textbf{[C2]} to be consistent &  \\ \cline{2-5}
\multicolumn{1}{|l|}{} & \multirow{2}{*}{\textit{Manners}} & \textbf{[B1]} to increase human-likeness & \textbf{[C1]} to deal with face-threatening acts & \textbf{[S1]} to engage in small talk \\
\multicolumn{1}{|l|}{} &  &  & \textbf{[C2]} to end a conversation gracefully & \textbf{[S2]} to adhere turn-taking protocols \\ \cline{2-5}
\multicolumn{1}{|l|}{} & \multirow{3}{*}{\textit{Moral agency}} & \textbf{[B1]} to avoid stereotyping & \textbf{[C1]} to avoid alienation & Not identified \\
\multicolumn{1}{|l|}{} &  & \textbf{[B2]} to enrich interpersonal relationships & \textbf{[C2]} to build unbiased training data and algorithms &  \\ \cline{2-5}
\multicolumn{1}{|l|}{} & \multirow{3}{*}{\textit{\shortstack{Emotional\\intelligence}}} & \textbf{[B1]} to enrich interpersonal relationships & \textbf{[C1]} to regulate affective reactions & \textbf{[S1]} to use social-emotional utterances \\
\multicolumn{1}{|l|}{} &  & \textbf{[B2]} to increase engagement &  & \textbf{[S2]} to manifest \textit{conscientiousness} \\
\multicolumn{1}{|l|}{} &  & \textbf{[B3]} to increase believability &  & \textbf{[S3]} reciprocity and self-disclosure \\ \cline{2-5}
\multicolumn{1}{|l|}{} & \multirow{3}{*}{\textit{Personalization}} & \textbf{[B1]} to enrich interpersonal relationships & \textbf{[C1]} privacy & \textbf{[S1]} to learn from and about the user \\
\multicolumn{1}{|l|}{} &  & \textbf{[B2]} to provide unique services &  & \textbf{[S2]} to provide customizable agents \\
\multicolumn{1}{|l|}{} &  & \textbf{[B3]} to reduce interactional breakdowns &  & \textbf{[S3]} visual elements \\ \hline
\multicolumn{1}{|l|}{\multirow{5}{*}{\rotatebox{90}{\textbf{\shortstack{Personifi-\\cation}}}} }& \multirow{3}{*}{\textit{Identity}} & \textbf{[B1]} to increase engagement & \textbf{[C1]} to avoid negative stereotypes & \textbf{[S1]} to design and elaborate on a persona \\
\multicolumn{1}{|l|}{} &  & \textbf{[B2]} to increase human-likeness & \textbf{[C2]} to balance the \textit{identity} and the technical capabilities &  \\\cline{2-5}
\multicolumn{1}{|l|}{} & \multirow{2}{*}{\textit{Personality}} & \textbf{[B1]} to increase believability & \textbf{[C1]} to adapt humor to the users' culture & \textbf{[S1]} to use appropriate language \\
\multicolumn{1}{|l|}{} &  & \textbf{[B2]} to enrich interpersonal relationships & \textbf{[C2]} to balance the \textit{personality} traits & \textbf{[S2]} to have a sense of humor \\
\hline
\end{tabular}
\label{tab:conceptualmodel}
\end{table}
\end{landscape}

\section{Chatbots Social Characteristics}
\label{sec:socialcharacteristics}

This section describes the identified social characteristics grouped into categories. As Table \ref{tab:conceptualmodel} depicts, the category \textbf{conversational intelligence} includes characteristics that help the chatbot manage interactions. \textbf{Social intelligence} focuses on habitual social protocols, while \textbf{personification} refers to the chatbot's perceived identity and personality representations. We also grouped the social characteristics based on the domain in which they were investigated (Table \ref{tab:domain-analysis}). In the following subsections, we describe the identified social characteristics as well as the domains of study. Then, we summarize the relationship among the characteristics in Section \ref{sec:interrelationship}. For each category, a table with an overview of the surveyed studies is provided in the supplementary materials. The supplementary materials also include tables for each social characteristic, listing the studies associated with the domains of study and reported benefits, challenges, and strategies. Finally, the supplementary materials also highlight five constructs that can be used to assess whether social characteristics reach the intended design goals.

\subsection{Conversational Intelligence}
\label{sec:conversationalintelligence}

\textbf{Conversational intelligence} enables the chatbot to actively participate in the conversation and to demonstrate awareness of the topic discussed, the evolving conversational context, and the flow of the dialogue. Therefore, \textbf{conversational Intelligence} refers to the ability of a chatbot to effectively converse beyond the technical capability of achieving a conversational goal \citep{jain2018evaluating}. In this section, we discuss social characteristics related to \textbf{conversational intelligence}, namely: \textit{proactivity} (18 studies), \textit{conscientiousness} (11 studies), and \textit{communicability} (6 studies). Most of the studies rely on data that comes from the log of the conversations, interviews, and questionnaires. The questionnaires are mostly Likert-scales, and some of them include subjective feedback. Most studies analyzed the interaction with real chatbots, although Wizard of Oz (WoZ) settings are also common. Only two papers did not evaluate a particular type of interaction because they were based on a literature review \citep{gnewuch2017towards} and surveys with chatbot users in general \citep{brandtzaeg2017people}. Only five studies applied only quantitative methods, while seven focused on qualitative methods. The majority of the studies (15) applied mixed methods (both quantitative and qualitative). See the supplementary materials for details.

\begin{table}[!btp]
\scriptsize
\centering
\caption{Studied social characteristics per domain. Research in open domain chatbots has reported most of the social characteristics, except for \textit{Communicability}. In he task-oriented domains, some characteristics are largely influenced by the topic (e.g., \textit{Moral agency} and \textit{Personality}), while others are more generally applied (e.g., \textit{Manners} and \textit{Damage control}).}
\begin{tabular}{p{1.6cm}p{5cm}p{6.5cm}}
\hline
\textbf{Domain} & \textbf{Social Characteristics} & \textbf{Studies} \\
\hline
\multirow{2}{*}{\shortstack[l]{Open\\domain}} & \textit{Proactivity, Conscientiousness, Damage control, Thoroughness, Manners, Moral agency, Emotional intelligence, Personalization, Identity, Personality} & \citep{thies2017how} \citep{portela2017new} \citep{shum2018eliza} \citep{morrissey2013realness} \citep{curry2018metoo} \citep{deangeli2001unfriendly} \citep{hill2015real} \citep{kirakowski2009establishing} \citep{mairesse2009can} \citep{deangeli2006sex} \citep{banks2018perceived} \citep{brahnam2012gender} \citep{ho2018psychological} \citep{deangeli2005rescue} \citep{corti2016co} \citep{ptaszynski2010towards} \\ 
\hline
Ethnography & \textit{Proactivity, Conscientiousness, Thoroughness, Personalization} & \citep{tallyn2018ethnobot} \\ 
\hline
\multirow{2}{*}{\shortstack[l]{Task\\management}} & \textit{Proactivity, Damage control, Manners, Personalization, Identity} & \citep{liao2016what} \citep{toxtli2018understanding} \\ 
\hline
Tourism & \textit{Proactivity, Thoroughness, Manners} & \citep{chaves2018single} \\ 
\hline
Business & \textit{Proactivity, Personalization} & \citep{duijvelshoff2017use} \\ 
\hline
Information search & \textit{Proactivity, Damage control, Manners, Emotional intelligence} & \citep{avula2018searchbots} \citep{wallis2005trouble} \\ 
\hline
Decision-making & \textit{Proactivity, Damage control, Manners} & \citep{maurer2015benjamin} \\
\hline
Health-care & \textit{Proactivity, Emotional intelligence} & \citep{fitzpatrick2017delivering} \citep{miner2016conversational} \\ 
\hline
\shortstack{Credibility\\assessment} & \textit{Proactivity, Conscientiousness} & \citep{schuetzler2018investigation} \\ 
\hline
Education & \textit{Proactivity, Conscientiousness, Damage control, Thoroughness, Manners, Emotional intelligence, Identity, Personality} & \citep{ayedoun2017communication} \citep{coniam2008evaluating} \citep{dyke2013towards} \citep{hayashi2015social} \citep{kumar2010socially} \citep{silvervarg2013iterative} \citep{sjoden2011extending} \citep{tamayo2016adapting} \citep{tegos2016investigation} \\ 
\hline
\multirow{2}{*}{\shortstack[l]{Financial\\services}} & \textit{Conscientiousness, Communicability, Damage control, Thoroughness, Personalization, Identity} & \citep{candello2017typefaces} \citep{duijst2017can} \\ 
\hline
\multirow{2}{*}{\shortstack[l]{Customer\\services}} & \textit{Conscientiousness, Communicability, Damage control, Thoroughness, Manners, Emotional intelligence, Personalization, Identity} & \citep{araujo2018living} \citep{brandtzaeg2018chatbots} \citep{gnewuch2017towards} \citep{jenkins2007analysis} \citep{lasek2013chatbots} \\ 
\hline
E-commerce & \textit{Conscientiousness, Manners} & \citep{jain2018convey} \citep{narita2010persuasive} \\ 
\hline
News & \textit{Communicability} & \citep{valerio2017here} \\ 
\hline
\multirow{2}{*}{\shortstack[l]{Human\\resources}} & \textit{Communicability, Damage control, Manners, Identity} & \citep{liao2018all} \\ 
\hline
\multirow{2}{*}{\shortstack[l]{Virtual\\assistant}} & \textit{Thoroughness, Emotional intelligence, Personalization, Identity} & \citep{ciechanowski2018shades} \citep{zamora2017sorry} \\ 
\hline
Gaming & \textit{Thoroughness, Emotional intelligence, Personality} & \citep{dohsaka2014effects} \citep{morris2002conversational} \\ 
\hline
Race-talk & \textit{Moral agency, Identity} & \citep{marino2014racial} \citep{schlesinger2018let} \\ 
\hline
Humorous talk & \textit{Personality} & \citep{meany2010humour} \\ 
\hline
Not defined & \textit{Proactivity, Conscientiousness, Communicability, Damage control, Personalization, Identity, Personality} & \citep{brandtzaeg2017people} \citep{jain2018evaluating} \citep{neururer2018perceptions} \\
\hline
\end{tabular}
\label{tab:domain-analysis}
\end{table}

\subsubsection{Proactivity}
\label{sec:proactivity}

\textit{Proactivity} is the capability of a system to 
autonomously act on the user's behalf \citep{salovaara2004six} and thereby reduce the amount of human effort to complete a task \citep{tennenhouse2000proactive}. In human-chatbot conversations, a proactive behavior enables a chatbot to share initiative with the user, contributing to the conversation in a more natural way \citep{morrissey2013realness}. Chatbots may manifest \textit{proactivity} when they initiate exchanges, suggests new topics, provide additional information, or formulate follow-up questions. In this survey, we found 18 papers that report either chatbots with proactive behavior or implications of manifesting a proactive behavior. \textit{Proactivity} (also addressed as ``\textit{intervention mode}'') was explicitly addressed in seven studies \citep{liao2016what, avula2018searchbots, schuetzler2018investigation, chaves2018single, hayashi2015social, tegos2016investigation, dyke2013towards}. In most of the studies, however, \textit{proactivity} emerged either as an exploratory result, mostly from post-intervention interviews and user's feedback \citep{portela2017new, shum2018eliza, jain2018evaluating, duijvelshoff2017use, thies2017how, morrissey2013realness}, or as a strategy to attend to domain-specific requirements (e.g., monitoring, and guidance) \citep{silvervarg2013iterative, maurer2015benjamin, tallyn2018ethnobot, toxtli2018understanding, fitzpatrick2017delivering}. \textit{Proactivity} was mostly investigated in open domain and education chatbots (four studies each). In open domain chatbots, \textit{proactivity} helps improve engagement by introducing new topics to keep the conversation alive \citep{shum2018eliza}. Educational chatbots rely on \textit{proactivity} to prompt students to think, share, and collaborate (e.g., \citet{dyke2013towards}). \textit{Proactivity} was also observed in eight other task-oriented domains, including task management and information searches as can be observed in the supplementary materials.

The surveyed literature evidences several benefits of chatbot \textit{proactivity} in chatbots: 

\textbf{[B1] to provide additional, useful information:} literature reveals that \textit{proactivity} in chatbots adds value to interactions \citep{morrissey2013realness, thies2017how, avula2018searchbots}. Investigating evaluation criteria for chatbots, \citet{morrissey2013realness} asked users of a general purpose chatbot to rate the chatbots' naturalness and report in what areas they excel. Both statistical and qualitative results confirm that taking the lead and suggesting specialized information about the conversation theme correlate to chatbots' naturalness. \citet{thies2017how} corroborates this result; in post-intervention interviews, ten out of 14 users mentioned they preferred a chatbot that takes the lead and volunteers additional information, such as useful links and song playlists. In a WoZ study, \citet{avula2018searchbots} investigated whether proactive interventions of a chatbot contribute to a collaborative search in a group chat. The chatbot either elicits or infers needed information from the collaborative chat and proactively intervenes in the conversation by sharing useful search results. The intervention modes were not significantly different from each other, but both intervention modes resulted in a statistically significant increase of enjoyment and decrease of effort when compared to the same task with no chatbot interventions. Moreover, in a post-intervention, open-ended question, 16 out of 98 participants self-reported positive perceptions about the provided additional information.

\textbf{[B2] to inspire users, and keep the conversation alive: }proactively suggesting and encouraging new topics have been shown useful to both inspire users \citep{chaves2018single, avula2018searchbots} and keep the conversation alive \citep{silvervarg2013iterative}. Participants in the study conducted by \citet{avula2018searchbots} self-reported that the chatbot's suggestions helped them to get started (7 mentions) and gave them ideas about search topics (4 mentions). After iteratively evaluating prototypes for a chatbot in an educational scenario, \citet{silvervarg2013iterative} concluded that proactively initiating topics makes the dialogue more fun and reveals topics the chatbot can talk about. The refined prototype also proactively maintains the engagement by posing a follow-up when the student had not provided an answer to the question. \citet{schuetzler2018investigation} hypothesized that including follow-up questions based on the content of previous messages would result in higher perceived partner engagement. The hypothesis was supported, with participants in the dynamic condition rating the chatbot as more engaging. In an ethnographic data collection \citep{tallyn2018ethnobot}, users included photos in their responses to add information about their experience; 85\% of these photos were proactively prompted by the chatbot. This result shows that prompting the user for more information stimulates them to expand their entries. \citet{chaves2018single} also observed that chatbots' proactive messages provided insights about the chatbots' knowledge, which potentially helped the conversation to continue. In this paper, we call the strategies to convey the chatbot's knowledge and capabilities as \textit{communicability}, and we discuss it in Section \ref{sec:communicability}.

\textbf{[B3] to recover the chatbot from a failure: }in \citet{portela2017new} and \citet{silvervarg2013iterative}, \textit{proactivity} is employed to naturally recover from a failure. In both studies, the approach was to introduce a new topic when the chatbot failed to understand the user or could not find an answer, preventing the chatbot from getting stuck and keeping the conversation alive. Additionally, in \citet{silvervarg2013iterative}, the chatbot inserted new topics when users are either abusive or non-sensical. We call the strategies to handle failure and abusive behavior as \textit{damage control}, and we discuss this characteristic in Section \ref{sec:damagecontrol}.

\textbf{[B4] to improve conversation productivity: }in task-oriented interactions, such as searching or shopping, \textit{proactivity} can improve the conversation's productivity \citep{jain2018evaluating}. In interviews with first-time users of chatbots, \citet{jain2018evaluating} found that chatbots should ask follow-up questions to resolve and maintain the context of the conversation and reduce the time searching before achieving the goal. \citet{avula2018searchbots} found similar results for collaborative searches; 28 out of 98 participants self-reported that chatbot's proactive interventions saved collaborators time.

\textbf{[B5] to guide and engage users: }in particular domains, \textit{proactivity} helps chatbots to either guide users or establish and monitor users' goals. In \citet{fitzpatrick2017delivering}, the chatbot assigns a goal to the user and proactively prompts motivational messages and reminders to keep the user engaged in the treatment. \citet{maurer2015benjamin} suggest that a decision-making coach chatbot needs to lead the interaction toward guiding the user to a decision. In ethnographic data collection \citep{tallyn2018ethnobot}, the chatbot prompts proactive messages that guide the users on what information they need to report. \citet{toxtli2018understanding} evaluates a chatbot that manage tasks in a workplace. Proactive messages are used to check whether the team member has concluded the tasks, and then report the outcome to the other stakeholders. In the educational context, \textit{proactivity} is used to develop tutors that engage the students and facilitate learning. In \citet{hayashi2015social}, the tutor chatbot was designed to provide examples of how other students explained a topic. The network analysis of the learner's textual inputs shows that students used more key terms and provided more important messages when receiving feedback about other group members. In \citet{hayashi2015social}, \citep{dyke2013towards}, and \citep{tegos2016investigation} the chatbots prompt utterances to encourage the students to reason about a topic. In all three studies, the chatbot condition provided better learning outcomes and increased students' engagement in the discussions.

The surveyed papers also highlight challenges in providing proactive interactions, such as timing and relevance, privacy, and the user's perception of being controlled.

\textbf{[C1] timing and relevance:} untimely and irrelevant proactive messages may compromise the success of the interaction. \citet{portela2017new} states that untimely turn-taking behavior was perceived as annoying, negatively affecting emotional engagement. \citet{liao2016what} and \citet{chaves2018single} reported that \textit{proactivity} can be disruptive. \citet{liao2016what} investigated \textit{proactivity} in a workspace environment, hypothesizing that the perceived interruption of agent \textit{proactivity} negatively affects users' opinion's. The hypothesis was supported, and the authors found that influencing the sense of interruption was the general aversion to unsolicited messages, regardless of whether they came from a chatbot or a colleague. \citet{chaves2018single} showed that proactively introducing new topics resulted in a high number of ignored messages. The analysis of the conversation log reviewed that either the new topics were not relevant, or it was not the proper time to start a new topic. \citet{silvervarg2013iterative} also reported annoyance when a chatbot introduces repetitive topics.

\textbf{[C2] privacy:} in a work-related, group chat, \citet{duijvelshoff2017use} observed privacy concerns regarding the chatbot ``reading'' the employees' conversations to proactively act. During a semi-structured interview, researchers presented a mockup of the chatbot to employees from two different enterprises and collected perceptions of usefulness, intrusiveness, and privacy. Employees reported feeling that the chatbot represented their supervisors' interests, which conveyed as sense of workplace surveillance. Privacy concerns may result in under-motivated users, discomfort about disclosing information, and lack of engagement \citep{duijvelshoff2017use}.

\textbf{[C3] user's perception of being controlled:} \textit{proactivity} can be annoying when the chatbot conveys the impression of trying to control the user. \citet{tallyn2018ethnobot} report that seven out of 13 participants experienced irritation with the chatbot; one of the most frequent reasons was due to the chatbot directing them to specific places. For the task management chatbot, \citet{toxtli2018understanding} reported to have adapted the follow-up questions approach to pose questions in a time negotiated with the user. In a previous implementation, the chatbot checked the status of the task twice a day, which participants considered too frequent and annoying.

The surveyed literature also reveals two strategies to provide \textit{proactivity}: leveraging the conversational context and randomly selecting a topic. \textbf{[S1] Leveraging the conversational context} is the most frequent strategy \citep{chaves2018single, avula2018searchbots, shum2018eliza, duijvelshoff2017use, thies2017how}, in which proactive messages relate to contextual information provided in the conversation to increase the usefulness of interventions \citep{avula2018searchbots, duijvelshoff2017use, shum2018eliza}. \citet{shum2018eliza} argue that general purpose, emotionally aware chatbots should recognize users' interests and intents from the conversational context to proactively offer comfort and relevant services. In \citet{duijvelshoff2017use}, the chatbot leverages conversational context to suggest new topics and propose to share documents or links to assist employees in a work-related group chat. The chatbots studied by \citet{chaves2018single} introduce new topics based on keywords from previous utterances posted in the chat. According to \citet{shum2018eliza}, leveraging the context can also help smoothly guide the user to a target topic. Researchers in the chatbots domain can refer to the emerging literature on Ambient Intelligence (see, e.g., ~\citep{stefanidi2019parlami, leonidis2017using}) to understand how contextual knowledge can be leveraged to convey proactivity. One surveyed paper \citep{portela2017new} proposes a chatbot that \textbf{[S2] selects a topic randomly} but also observes that the lack of context is a major problem of this approach. Contextualized proactive interventions also suggest that the chatbot should be attentive to the conversation, which conveys \textit{conscientiousness}, as discussed in the next section.

\subsubsection{Conscientiousness}
\label{sec:conscientiousness}

\textit{Conscientiousness} is a chatbot's capacity to demonstrate attentiveness to the conversation at hand \citep{dyke2013towards, duijst2017can}. It enables a chatbot to follow the conversational flow, show understanding of the context, and interpret each utterance as a meaningful part of the whole conversation \citep{morrissey2013realness}. In this survey, we found 11 papers that reported findings related to \textit{conscientiousness} for chatbot design. Four studies explicitly investigated influences of \textit{conscientiousness} for chatbots \citep{jain2018convey, schuetzler2018investigation, coniam2008evaluating, ayedoun2017communication}. In the remaining studies, \textit{conscientiousness} emerged in exploratory findings. In \citet{dyke2013towards}, \textit{conscientiousness} emerged from the analysis of conversational logs, while \citet{gnewuch2017towards} elicited \textit{conscientiousness} aspects as a requirement for chatbot design when surveying the literature on customer service chatbots. In the remaining studies \citep{brandtzaeg2017people, jain2018evaluating, tallyn2018ethnobot, morrissey2013realness, duijst2017can}, \textit{conscientiousness} issues were self-reported by the users in post-intervention interviews and subjective feedback to open-ended questions. \textit{Conscientiousness} was investigated in three studies from the education domain, where the tutor chatbot is expected to be attentive and relevant as well as to control the conversation flow to keep the students focused on the topic (see e.g., \citet{dyke2013towards}). \textit{Conscientiousness} was also investigated in seven other domains, including open domain interactions and customer and financial services. See the supplementary materials for a complete list of domains of studies that report \textit{conscientiousness} as a social characteristic for chatbots.

The surveyed literature evidenced benefits of designing conscientious chatbots:

\textbf{[B1] to provide meaningful answers:} some chatbots use simplistic approaches, like pattern matching rules based on keywords or template phrases applied in the last user utterance, to find the most appropriate response \citep{abdul2015survey, ramesh2017survey}. However, as the chatbot does not interpret the meaning and the intent of users' utterances, the best-selected response may still sound irrelevant to the conversation \citep{dyke2013towards, duijst2017can, coniam2008evaluating}. As shown by \citet{duijst2017can}, when a chatbot does not interpret the meaning of users' utterances, users show frustration and the chatbot's credibility is compromised. This argument is supported by \citet{dyke2013towards} when studying chatbots to facilitate collaborative learning. The authors proposed a chatbot that promotes Academically Productive Talk moves. Exploratory results show that the chatbot performed inappropriate interventions, which was perceived as a lack of attention to the conversation. In \citet{jain2018evaluating}, participants complained that some chatbots seemed ``\textit{completely scripted,}'' ignoring user's inputs that did not fall into the script. In this case, users needed to adapt their inputs to match the chatbot script to be understood, which resulted in dissatisfaction. Besides avoiding frustration, \citet{schuetzler2018investigation} showed that \textit{conscientiousness} also influences chatbots' perceived humanness and social presence. The authors invited participants to interact with a chatbot that shows an image and asks the users to describe it. To perform the task, the chatbot could either ask the same generic follow-up questions each time (nonrelevant condition) or respond with a follow-up question related to the last participant's input (relevant condition), demonstrating attention to the information provided by the user. Statistical analysis of survey rates supported that the relevant condition increased chatbots' perceived humanness and social presence. In \citet{ayedoun2017communication}, to motivate the students to communicate in a second language, the proposed chatbot interprets users' input to detect breakdowns and react using an appropriate communication strategy. In interviews, participants reported ``\textit{appreciating the help they got from the chatbot to understand and express what they have got to say}'' \citep{ayedoun2017communication}. This study was later extended to the context of ECAs (see \citep{ayedoun2018adding} for details).

\textbf{[B2] to hold a continuous conversation: }a conversation with a chatbot should maintain a ``\textit{sense of continuity over time}'' \citep{jain2018evaluating} to demonstrate that the chatbot is exerting effort to track the conversation. To do so, it is essential to maintain the topic. When evaluating the naturalness of a chatbot, \citet{morrissey2013realness} found that maintaining a theme is convincing, while failure to do so is unconvincing. Furthermore, based on the literature on customer support chatbots, \citet{gnewuch2017towards} argue that comfortably conversing on any topic related to the service offering is a requirement for task-oriented chatbots. \citet{coniam2008evaluating} reviewed popular chatbots for practicing second languages. The author showed that most chatbots in this field cannot hold continuous conversations, since they are developed to answer the user's last input. Therefore, they did not have the sense of topic, which resulted in instances of inappropriate response. While the chatbots could change the topic, they could not sustain it afterward. Showing \textit{conscientiousness} also requires the chatbot to understand and track the context, which is particularly important in task-oriented scenarios. In \citet{jain2018evaluating}, first-time users stressed positive experiences with chatbots that retained information from previous turns. Two participants also expected the chatbots to retain this context across sessions, thus reducing the need for the user's additional input per interaction. Keeping the context across sessions was highlighted as a strategy to convey \textit{personalization} and empathy (see Sections \ref{sec:personalization} and \ref{sec:emotionalintelligence}).

\textbf{[B3] to steer the conversation toward a productive direction:} in task-oriented interactions, a chatbot should understand the purpose of the interaction and strive to conduct the conversation toward this goal in an efficient, productive way \citep{duijst2017can, ayedoun2017communication}. \citet{brandtzaeg2017people} show that productivity is the key motivating factor for using chatbots (68\% of the participants mentioned it as the main reason for using chatbots). First-time users in \citet{jain2018evaluating} self-reported that interacting with chatbots should be more productive than using websites, phone apps, and search engines. In this sense, \citet{duijst2017can} compared the user experience when interacting with a chatbot for solving either simple or complex tasks in a financial context. The authors found that, for complex tasks, to keep the conversation on track the user must be aware of the next steps or why something is happening. In the educational context, \citet{ayedoun2017communication} proposed a dialogue management approach based on communication strategies to enrich a chatbot with the capability to express its meaning when faced with challenges. Statistical results show that the communication strategies, combined with affective backchannel (which is detailed in Section \ref{sec:emotionalintelligence}), are effective in motivating students to communicate and maintain the task flow. Thirty-two participants out of 40 reported that they preferred to interact with a chatbot with these characteristics. Noticeably, the chatbot's attentiveness to the interactional goal may not be evident to the user if the chatbot passively waits for the user to control the interaction. Thus, \textit{conscientiousness} relates to proactive ability, as discussed in Section \ref{sec:proactivity}. 

Nevertheless, challenges in designing conscientious chatbots are also evident in the literature:

\textbf{[C1] to handle task complexity: }as the complexity of tasks increases, more turns are required to achieve a goal; hence, more mistakes may be made. This argument was supported by both \citet{duijst2017can} and \citet{dyke2013towards}, where the complexity of the task compromised the experience and satisfaction in using the chatbot. \citet{duijst2017can} also highlights that complex tasks require more effort to correct eventual mistakes. Therefore, it is an open challenge to design flexible workflows, where the chatbot recovers from failures and keeps the interaction moving productively toward the goal, despite potential misunderstandings \citep{gnewuch2017towards}. 
Recovering from failure 
is discussed in Section \ref{sec:damagecontrol}.

\textbf{[C2] to harden the conversation:} aiming to ensure the conversational structure--and to hide natural language limitations--chatbots are designed to restrict free-text inputs from the user \citep{duijst2017can, jain2018evaluating, tallyn2018ethnobot}. However, limiting the choices of interaction may convey a lack of attention to the users' inputs. In \citet{duijst2017can}, one participant mentioned the feeling of ``\textit{going through a form or a fixed menu.}'' According to \citet{jain2018evaluating}, participants consider the chatbot's understanding of free-text input as a criterion to determine whether it can be considered a chatbot, since chatbots are supposed to chat. In the ethnographic data collection study, \citet{tallyn2018ethnobot} reported that eight out of ten participants described the interaction using pre-set responses as too restrictive, although they fulfilled the purpose of nudging participants to report their activities. Thus, the challenge lies in how to leverage the benefits of suggesting predefined inputs without limiting conversational capabilities.

\textbf{[C3] to keep the user aware of the chatbot's context:} a chatbot should provide a way to inform the user of the current context, especially for complex tasks. According to \citet{jain2018convey}, context can be inferred from explicit user input, or assumed based on data from previous interactions. In both cases, the user and chatbot should be on the same page about the chatbot's contextual state \citep{jain2018convey}, providing the users the opportunity to clarify possible misunderstandings \citep{gnewuch2017towards}. \citet{jain2018evaluating} highlighted that participants reported negative experience when finding ``mismatch[es] between [a] chatbot's real context and their assumptions of the chatbot context.'' We identified three strategies used to provide understanding to a chatbot:

\textbf{[S1] conversation workflow: }designing a conversational blueprint helps to conduct the conversation strictly and productively to the goal \citep{duijst2017can}. However, \citet{gnewuch2017towards} argue that the workflow should be flexible enough to handle both multi-turn and one-turn question-answer interactions; it also should be unambiguous, such that users can efficiently achieve their goals. In addition, \citet{duijst2017can} discuss that the workflow should make it easy to fix mistakes; otherwise, the users need to restart the workflow, which leads to frustration. In \citet{ayedoun2017communication}, the conversation workflow included communicative strategies to detect a learner's breakdowns and pitfalls. In that study, when the student does not respond, the chatbot uses a comprehension-check question to detect whether the student understood what was said. Then, it reacts to the user's input by adopting one of the proposed communication strategies (e.g., asking for repetition or simplifying the previous sentence). The conversation workflow could also allow the chatbot to be proactive
. For example, participants in \citet{jain2018evaluating} suggested that proactive follow-up questions would anticipate the resolution of the context, reducing the effort required from the user to achieve the goal.

\textbf{[S2] visual elements: }user-interface resources--such as quick replies, cards, and carousels
--are used to structure the conversation and reduce issues regarding understanding \citep{duijst2017can, jain2018evaluating, tallyn2018ethnobot}. Using these resources, the chatbot shows the next possible utterances \citep{jain2018evaluating} and conveys the conversational workflow in a step-by-step manner \citep{duijst2017can, tallyn2018ethnobot}. Visual elements are also used to show the user what the chatbot can (or cannot) do. This is another conversational characteristic, which will be discussed in the 
Section \ref{sec:communicability}.

\textbf{[S3] context window:} to keep the user aware of the chatbot's current context, \citet{jain2018convey} developed a chatbot for shopping that shows a context window on the side of the conversation. In this window, the user can click on specific attributes and change them to fix inconsistencies. A survey showed that the chatbot outperformed a default chatbot (without the context window) for mental demand and effort constructs. However, when the chatbots are built into third-party apps (e.g., Facebook Messenger), an extra window may not be possible.

\textbf{[S4] confirmation messages: }a conversation workflow may include confirmation messages to convey the chatbots' context to the user \citep{jain2018convey}. In \citet{duijst2017can}, when trying to block a stolen credit card, a confirmation message is used to verify the given personal data. In \citet{ayedoun2017communication}, confirmation messages are used as a communicative strategy to check whether the system's understanding about a particular utterance matches what the user actually meant. Balancing the number of confirmation messages (see \citep{duijst2017can}) and the right moment to introduce them into the conversation flow is still under-investigated.

The surveyed literature supports that failing to demonstrate understanding about the users' individual utterances, the conversational context, and the interactional goals results in frustration and loss of credibility. However, most of the results are exploratory; a lack of studies investigate the extent to which the provided strategies influence users' behaviors and perceptions. In the field of human-human communication, the cooperative principle \citep{grice1975logic} has been extensively applied to understand conversational partners' expectations during a conversation. Additionally, recent studies have shown that this principle influences chatbots' perceived humanness \citep{jacquet2019impact, jacquet2018gricean}. Particularly, the maxims of relation (the ability to be relevant) and manner (the ability to be clear and orderly) can contribute to the design of \textit{conscientious} chatbots. In addition, \textit{conscientiousness} is by itself a \textit{personality} trait; the more \textit{conscientiousness} a chatbot manifests, the more it can be perceived as attentive, organized, and efficient. The relationship between \textit{conscientiousness} and \textit{personality} is highlighted in Section \ref{sec:interrelationship}.

\subsubsection{Communicability}
\label{sec:communicability}

Interactive software is communicative by its nature, since users achieve their goals 
by exchanging messages with the system \citep{sundar2016theoretical, prates2000methods}. In this context, \textit{communicability} is defined as the capacity of a software to convey to users its underlying design intent and interactive principles \citep{prates2000methods}. Providing \textit{communicability} helps users to interpret the codes used by designers to convey the interactional possibilities embedded in the software \citep{desouza1999method}, which improves system learnability \citep{grossman2009survey}. In the chatbot context, \textit{communicability} is, therefore, the capability of a chatbot to convey its features to users \citep{valerio2017here}. The problema around chatbots' \textit{communicability} lies in the nature of the interface: instead of buttons, menus, and links, chatbots unveil their capabilities through conversational turns, one sentence at a time \citep{valerio2017here}, bringing new challenges to the system learnability field. The lack of \textit{communicability} may lead users to give up on using the chatbot when they cannot understand the available functionalities and how to use them \citep{valerio2017here}.

In this survey, we found six papers that describe \textit{communicability}, although investigating \textit{communicability} is the main purpose of only one \citep{valerio2017here}. Conversational logs revealed \textit{communicability} needs in three studies \citep{liao2018all, lasek2013chatbots}; in the other three studies \citep{jain2018evaluating, gnewuch2017towards, duijst2017can} \textit{communicability} issues were self-reported by the users in post-intervention interviews and subjective feedback. \textit{Communicability} was investigated in the customer services domain (two studies), where chatbots guide customers through available functionalities \citep{valerio2017here}. The remaining task-oriented domains in which \textit{communicability} was investigated are financial services, news, and human resources. We did not identify studies on open domain chatbots that report on \textit{communicability}, since in open domain interactions users are free to talk about varying topics and guidance is less a concern.

The surveyed literature reports two main benefits of \textit{communicability} for chatbots:

\textbf{[B1] to unveil functionalities: }while interacting with chatbots, users may not know that a desired functionality is available or how to use it \citep{jain2018evaluating, valerio2017here}. Most participants in \citet{jain2018evaluating}'s study mentioned that they did not understand the functionalities of at least one of the chatbots and none of them mentioned searching for the functionalities in other sources (e.g., Google search or the chatbot website) rather than exploring options during the interaction. In a study about playful interactions in a work environment, \citet{liao2018all} observed that 22\% of the participants explicitly asked the chatbot about its capabilities (e.g., \textit{``what can you do?''}), and 1.8\% of all the users' messages were ability-check questions. In a study about hotel chatbots, \citet{lasek2013chatbots} verified that 63\% of the conversations were initiated by clicking an option displayed in the chatbot welcome message. A semiotic inspection of news-related chatbots \citep{valerio2017here} evidenced that \textit{communicability} strategies are effective at providing clues about the chatbot's features and ideas about what to do and how.

\textbf{[B2] to manage users' expectations:} \citet{jain2018evaluating} observed that when first-time users do not understand chatbots' capabilities and limitations, they have high expectations and, consequently, end up more frustrated when the chatbots fail. Some participants 
blamed themselves for not knowing how to communicate and gave up. 
In \citet{liao2018all}, quantitative results 
evidenced that ability-check questions can be considered signals of users struggling with functional affordances. Users posed ability-check questions after encountering errors as a means of establishing a common ground between the chatbot's capabilities and their own expectations. According to the authors, ability-check questions helped users to understand the system and reduce uncertainty \citep{liao2018all}. In \citet{duijst2017can}, users also demonstrated the importance of understanding chatbots' capabilities in advance.  Since the tasks related to financial support, users expected the chatbot to validate the personal data provided and to provide feedback after completing the task (e.g., explaining how long it would take for the credit card to be blocked). Therefore, \textit{communicability} helps users gain a sense of which type of messages or functionalities a chatbot can handle.

The surveyed literature also highlights two challenges of providing \textit{communicability}:

\textbf{[C1] to provide business integration: }communicating chatbots' functionalities should be performed as much as possible within the chat interface \citep{jain2018evaluating}. Chatbots often act as an intermediary between users and services. In this case, to overcome technical challenges, chatbots answer users' inputs with links to external sources, where the request will be addressed. First-time users expressed dissatisfaction with this strategy in \citet{jain2018evaluating}. Six participants complained that the chatbot redirected them to external websites. According to \citet{gnewuch2017towards}, business integration is a requirement for designing chatbots, so that the chatbot can solve the users' requests without transferring the interaction to another user interface.

\textbf{[C2] to keep visual elements consistent with textual inputs:} in the semiotic engineering evaluation, \citet{valerio2017here} observed that some chatbots responded differently depending on whether the user accesses a visual element in the user-interface or types the desired functionality in the text-input box, even if both input modes result in the same utterance. This inconsistency produces the user's feeling of misinterpreting the affordances, which has a negative impact on the system's learnability.

As an outcome of the semiotic inspection process, \citet{valerio2017here} present a set of strategies to provide \textit{communicability}. Some of them are also emphasized in other studies, as follows:

\textbf{[S1] to clarify the purpose of the chatbot:} First-time users in \citet{jain2018evaluating} highlighted that a clarification about the chatbots' purpose should be placed in the introductory message. \citet{gnewuch2017towards} found similar inference from the literature on customer services chatbots. The authors argue that providing an opening message with insights into the chatbots' capabilities is a requirement for chatbots design. In addition, a chatbot could give a short tour throughout the main functionalities at the beginning of the first sessions \citep{valerio2017here}.

\textbf{[S2] to advertise the functionality and suggest the next step: }when the chatbot is not able to answer the user, or when it notices that the user is silent, it may suggest available features to stimulate the user to engage 
\citep{jain2018evaluating, valerio2017here}. In \citet{jain2018evaluating}, six participants mentioned that they appreciated when the chatbot suggested responses, for example by saying \textit{``try a few of these commands: ...''} \citep{jain2018evaluating}. \citet{valerio2017here} shows that chatbots use visual elements, such as cards, carousel, and menus (persistent or not) to show contextualized clues about the next answer, which both fulfills \textit{communicability} purpose and spares users from having to type.

\textbf{[S3] to provide a help functionality: }chatbots should recognize a \textit{``help''} input from the user, so it can provide instructions on how to proceed \citep{valerio2017here}. \citet{jain2018evaluating} reported that users highlighted this functionality as useful for the reviewed chatbots. Also, results from \citep{liao2018all} show that chatbots should be able to answer ability-check questions (e.g., ``\textit{what can you do?}'' or ``\textit{can you do [functionality]?}'').

The literature states the importance of communicating chatbots' functionality to the success of the interaction. Failing to provide \textit{communicability} 
leads the users to frustration and they often give up when they do not know how to proceed. The literature on interactive systems has highlighted system learnability as the most fundamental component for usability \citep{grossman2009survey}, and an easy-to-learn system should lead the user to perform well, even during their initial interactions. Thus, researchers in the chatbots domain can leverage the vast literature on system learnability to identify metrics and evaluation methodologies, as well as propose new forms of \textit{communicability} strategies that reduce the learnability issues in chatbot interactions. \textit{Communicability} may also be used as a strategy to avoid mistakes (\textit{damage control}), which will be discussed in Section \ref{sec:damagecontrol}.

\vspace{-1.5mm}
\begin{framed} \small
\vspace{-2mm}
In summary, the \textbf{conversational intelligence }category includes characteristics that help a chatbot to perform a proactive, attentive, and informative role in the interaction. Acknowledging that functional aspects play a crucial role in the ability of a chatbot to converse intelligently, the survey focuses on the characteristics that go beyond the functional aspects, looking through the lens of social perceptions. The highlighted benefits relate to how a chatbot manages the conversation to make it productive, interesting, and neat. To achieve that, designers and researchers should care for the timing and relevance of provided information, privacy, interactional flexibility, and consistency.
\vspace{-2mm}
\end{framed}
\vspace{-2mm}

\subsection{Social Intelligence}
\label{sec:socialintelligence}

\textbf{Social Intelligence} refers to the ability of an individual to produce adequate social behavior for the purpose of achieving desired goals \citep{bjorkqvist2000social}. In the HCI domain, the Media Equation theory \citep{reeves1996people} posits that people react to computers as social actors. Hence, when developing chatbots, it is necessary to account for the socially acceptable protocols for conversational interactions \citep{wallis2005trouble, walker2009endowing}. Chatbots should be able to respond to social cues during the conversation, accept differences, and manage conflicts \citep{salovey1990emotional} as well as be empathic and demonstrate caring \citep{bjorkqvist2000social}, which ultimately increase chatbots' authenticity \citep{neururer2018perceptions}. In this section, we discuss the social characteristics related to \textbf{social intelligence}, namely: \textit{damage control} (12 papers), \textit{thoroughness} (13 papers), \textit{manners} (10 papers), \textit{moral agency} (6 papers),\textit{ emotional intelligence} (14 papers), and \textit{personalization} (11 papers). Although the focus of the investigations is diverse, we found more studies where the focus of the investigation relates to a specific social characteristic, particularly \textit{moral agency} and \textit{emotional intelligence}, when compared to the \textbf{conversational intelligence} category. Regarding the adopted research methods, 12 studies applied qualitative methods and 10 focuses on quantitative methods. 16 studies reported both qualitative and quantitative results.

\subsubsection{Damage control}
\label{sec:damagecontrol}

\textit{Damage control} is the ability of a chatbot to deal with either conflict or failure situations. Although the Media Equation theory argues that humans socially respond to computers as they respond to other people \citep{reeves1996people}, the literature has shown that interactions with conversational agents are not quite equal to human-human interactions \citep{luger2016like, mou2017media, shechtman2003media}. When talking to a chatbot, humans are more likely to harass \citep{hill2015real}, test the agent's capabilities and knowledge \citep{wallis2005trouble}, and feel disappointed with mistakes \citep{maurer2015benjamin, jain2018evaluating}. When a chatbot does not respond appropriately, it may encourage the abusive behavior \citep{curry2018metoo} or disappoint the user \citep{jain2018evaluating, maurer2015benjamin}, which ultimately leads the conversation to fail \citep{jain2018evaluating}. Thus, it is necessary to enrich chatbots with the ability to recover from failures and handle inappropriate talk in a socially acceptable manner \citep{wallis2005trouble, jain2018evaluating, silvervarg2013iterative}.

In this survey, we found 12 studies that discuss \textit{damage control} as a relevant characteristic for chatbots, two of which focus on conflict situations, such as testing and flaming \citep{wallis2005trouble, silvervarg2013iterative}. In the remaining studies \citep{liao2018all, duijst2017can, toxtli2018understanding, maurer2015benjamin, lasek2013chatbots, curry2018metoo, gnewuch2017towards, jain2018evaluating, jenkins2007analysis, deangeli2001unfriendly}, needs for \textit{damage control} emerged from the analysis of conversational logs and users' feedback. \textit{Damage control} was mostly investigated for open domain (two studies) and customer service chatbots (three studies). In open domain interactions, users are free to wander among topics and testing or flaming tends to be more frequent \citep{hill2015real}. In the customer services context, the chatbot needs to avoid disappointing the user, as frustration may negatively reflect the business that the agent represents \citep{araujo2018living}. \textit{Damage control} was also identified in other six domains (see the supplementary materials), such as task management and financial services.

The surveyed literature highlights the following benefits of providing \textit{damage control} in chatbots:

\textbf{[B1] to appropriately respond to harassment:} chatbots are more likely to targets of profanity than humans would \citep{hill2015real}. When analyzing conversation logs from hotel chatbots, \citet{lasek2013chatbots} observed that 4\% of the conversations contained vulgar, indecent, and insulting vocabulary, and 2.8\% of all statements were abusive. Qualitative evaluation reveals that the longer the conversations last, the more users are encouraged to go beyond the chatbots main functions. In addition, sexual expressions represented 1.8\% of all statements. A similar number was found in a study with general purpose chatbots \citep{curry2018metoo}. When analyzing a corpus from the Amazon Alexa Prize 2017, the researchers estimated that about 4\% of the conversations included sexually explicit utterances. \citet{curry2018metoo} used utterances from this corpus to harass a set of state-of-art chatbots and analyze the responses. The results show that chatbots respond to harassment in a variety of ways, including nonsensical, negative, and positive responses. However, the authors highlight that the responses should align with the chatbot's goal to avoid encouraging the behavior or reinforcing stereotypes.

\textbf{[B2] to deal with testing:} abusive behavior is often used to test chatbots' social reactions \citep{wallis2005trouble, lasek2013chatbots}. During the evaluation of a virtual guide to the university campus \citep{wallis2005trouble}, a participant answered the chatbot's introductory greeting with ``\textit{moron,}'' likely hoping to see how the chatbot would answer. \citet{wallis2005trouble} argue that handling this type of testing helps the chatbots to establish limits and resolve social positioning. Other forms of testing were highlighted in \citet{silvervarg2013iterative}, including sending random letters, repeated greetings, laughs and acknowledgments, and posing comments and questions about the chatbot's intellectual capabilities. When analyzing conversations with a task management chatbot, \citet{liao2018all} observed that casually testing the chatbots' ``intelligence'' is a manifestation of satisfaction seeking. In \citet{jain2018evaluating}, first-time users appreciated when the chatbot successfully performed tasks when the user expected the chatbot to fail, which shows that satisfaction is influenced by the ability to provide a clever response when the user tests the chatbot.

\textbf{[B3] to deal with lack of knowledge: }chatbots often fail in a conversation due to lack of either linguistic or world knowledge \citep{wallis2005trouble}. \textit{Damage control} enables the chatbot to admit the lack of knowledge or cover up cleverly \citep{jain2018evaluating}. When analyzing the log of a task management chatbot, \citet{toxtli2018understanding} found out that the chatbot failed to answer 10\% of the exchanged messages. The authors suggest that the chatbot should be designed to handle novel scenarios when the current knowledge is not enough to answer the requests. In some task-oriented chatbots, though, the failure may not be caused by a novel scenario, but by an off-topic utterance. In the educational context, \citet{silvervarg2013iterative} observed that students posted off-topic utterances when they did not know what topics they could talk about, which led the chatbot to fail rather than help the users to understand its knowledge. In task-oriented scenarios, the lack of linguistic knowledge may lead the chatbots to get lost in the conversational workflow \citep{gnewuch2017towards}, compromising the success of the interaction. \citet{maurer2015benjamin} demonstrated that dialogue-reference errors (e.g., user's attempt to correct a previous answer or jumping back to an earlier question) are one of the major reasons for failed dialogues and they mostly result from chatbots' misunderstandings.

The literature also reveals some challenges to provide \textit{damage control}:

\textbf{[C1] to deal with unfriendly users:} \citet{silvervarg2013iterative} argue that users who want to test and find the system's borders are likely to never have a meaningful conversation with the chatbot no matter how sophisticated it is. Thus, there is a limit to which \textit{damage control} strategies will be effective to avoid testing and abuse. In \citet{deangeli2001unfriendly}, the authors observed human tendencies to dominate, be rude, and infer stupidity, which they call ``\textit{unfriendly partners.}'' After an intervention where users interacted with a chatbot for decision-making coaching, \citet{maurer2015benjamin} evaluated participants' self-perceived work and cooperation with the system. The qualitative results show that cooperative users are significantly more likely to give a higher rating for overall evaluation and decision efficiency. The qualitative analysis of the conversation log reveals that a few interactions failed because the users' motivations were curiosity and mischief rather than trying to solve the decision problem.

\textbf{[C2] to identify abusive utterances: }several chatbots are trained on ``clean'' data. Because they do not understand profanity or abuse, they may not recognize a statement as harassment, which makes it difficult to adopt answering strategies \citep{curry2018metoo}. \citet{curry2018metoo} shows that data-driven chatbots often provide non-coherent responses to harassment. Sometimes, these responses conveyed the impression of flirtatious or counter-aggression. Providing means to identify an abusive utterance is important to adopting \textit{damage control} strategies.

\textbf{[C3] to fit the response to the context:} \citet{wallis2005trouble} argue that humans negotiate conflict and social positioning well before reaching abuse. In human-chatbot interactions, however, predicting users' behavior toward the chatbots in a specific context to develop the appropriate behavior is a challenge. \textit{Damage control} strategies need to be adapted to both the social situation and the intensity of the conflict. For example, \citet{curry2018metoo} showed that being evasive about sexual statements may convey the impression of flirtatiousness, which would not be an acceptable behavior for a customer assistant or a tutor chatbot. In contrast, adult chatbots are supposed to flirt, so encouraging behaviors are expected in some situations. \citet{wallis2005trouble} argue that when the chatbot is not accepted as part of the social group it represents, it is discredited by the user, leading the interaction to fail. In addition, designing chatbots with too strong of reactions may lead to ethical concerns \citep{wallis2005trouble}. For \citet{bjorkqvist2000social}, choosing between peaceful or aggressive reactions in conflict situations is optional for socially intelligent individuals. Enriching chatbots with the ability to choose between the options is a challenge.

\textit{Damage control} strategies depend on the type of failure and the target benefit, as following:

\textbf{[S1] emotional reactions:} \citet{wallis2005trouble} suggest that when faced with abuse, a chatbot could be seen to take offense and respond in kind or act hurt. The authors argue that humans might feel inhibited about hurting the pretended feelings of a machine if the machine is willing to hurt a human's feelings too \citep{wallis2005trouble}. If escalating the aggressive behavior is not appropriate, the chatbot could withdraw from the conversation \citep{wallis2005trouble} to demonstrate that the user's behavior is not acceptable. In \citet{deangeli2001unfriendly}, the authors discuss that users appeared to be uncomfortable and annoyed whenever the chatbot pointed out any defect in the user or reacted to aggression, as this behavior conflicted with the user's perceived power relations. This strategy is also applied in \citet{silvervarg2013iterative}, where abusive behavior may lead the chatbot to stop responding until the student changes the topic. \citet{curry2018metoo} categorized responses from state-of-the-art conversational systems in a pool of emotional reactions, both positive and negative. The reactions include humorous responses, chastising and retaliation, and evasive responses, as well as flirtation and play-along utterances. To provide an emotional reaction, \textit{emotional intelligence} is also required. This category is presented in Section \ref{sec:emotionalintelligence}.

\textbf{[S2] authoritative reactions: }when facing testing or abuse, chatbots can communicate consequences \citep{silvervarg2013iterative} 
or call on the authority of others \citep{wallis2005trouble, toxtli2018understanding}. In \citet{wallis2005trouble}, although the wizard acting as a chatbot was conscientiously working as a campus guide, she answered a bogus caller with \textit{``This is the University of Melbourne. Sorry, how can I help you?''} The authors suggest that the wizard was calling on the authority of the university to handle the conflict, where being part of a recognized institution places the chatbot in a stronger social group. In \citet{silvervarg2013iterative}, when students recurrently harass the chatbot, the chatbot informs the student that further abuse will be reported to the (human) teacher (although the paper does not clarify whether the problem is, in fact, escalated to a human). \citet{toxtli2018understanding} and \citet{jenkins2007analysis} also suggest that chatbots could redirect users' problematic requests to a human attendant in order to avoid conflict situations.

\textbf{[S3] to ignore the user's utterance and change the topic:} \citet{wallis2005trouble} argue that ignoring abuse and testing is not a good strategy because it could encourage more extreme behaviors. It also positions the chatbot as an inferior individual, which is particularly harmful in scenarios where the chatbot should demonstrate a more prominent or authoritative social role (e.g., a tutor). However, this strategy has been found in some studies to handle lack of knowledge. When iteratively developing a chatbot for an educational context, \citet{silvervarg2013iterative} proposed initiating a new topic in one out of four user's utterances that the chatbot did not understand.

\textbf{[S4] \textit{conscientiousness} and \textit{communicability}:} successfully implementing \textit{conscientiousness} and \textit{communicability} may prevent errors; hence, strategies to provide these characteristics can also be used for \textit{damage control}. In \citet{silvervarg2013iterative}, when users utter out-of-scope statements, the chatbot could make it clear what topics are appropriate to the situation. For task-oriented scenarios, where the conversation should evolve toward a goal, \citet{wallis2005trouble} argue that the chatbot can clarify the purpose of the offered service when facing abusive behavior, bringing the user back to the task. \citet{jain2018evaluating} showed that describing chatbot's capabilities after failures in the dialogue was appreciated by first-time users. In situations where the conversational workflow is susceptible to failure, \citet{gnewuch2017towards} discuss that posting confirmation messages avoids trapping the users in the wrong conversation path. Participants in \citet{duijst2017can} also suggested back buttons as a strategy to fix mistakes in the workflow. In addition, the exploratory results about the user interface showed that having visual elements such as quick replies prevents errors, since they keep the users aware of what to ask and the chatbot is more likely to know how to respond \citep{duijst2017can}.

\textbf{[S5] to predict users' satisfaction:} chatbots should perceive both explicit and implicit feedback about users' (dis)satisfaction \citep{liao2018all}.
To address this challenge, \citet{liao2018all} invited participants to send a ``\textit{\#fail}'' statement to express dissatisfaction. The results show that 42.4\% of the users did it at least once, and the number of complaints and flaming for the proposed chatbot was significantly lower than the baseline. However, the amount of implicit feedback was also significant, which advocates for predicting user's satisfaction from the conversation. The most powerful conversational act that predicted user satisfaction in that study was the agent ability-check questions, see discussion in \textit{Communicability} section) and the explicit feedback \textit{\#fail}, although closings and off-topic requests were also significant in predicting frustration. Although these results are promising, more investigation is needed to identify other potential predictors of users' satisfaction in real-time, in order to provide appropriate reactions.

\textit{Damage control} strategies have different levels of severity. Deciding what strategy is adequate to the intensity of the conflict is crucial \citep{wallis2005trouble}. The strategies can escalate in severity if the conflict is not solved. For example, \citet{silvervarg2013iterative} uses a sequence of clarification, suggesting a new topic, and asking a question about the new topic. In case of abuse, the chatbot refers to an authority after two attempts at changing the topic.

According to \citet{wallis2005trouble}, humans also fail in conversations; they misunderstand what their partner says and do not know things that are assumed common knowledge by others. Hence, it is unlikely that chatbots interactions will evolve to be conflict-free. That said, \textit{damage control} intends to avoid escalating the conflicts and manifesting an unexpected behavior. In this sense, politeness can be used as a strategy to minimize the effect of lack of knowledge (see Section \ref{sec:manners}), managing the conversation despite the possible mistakes. Regarding interpersonal conflicts, the strategies are in line with the theory on human-human communication, which includes non-negotiation, emotional appeal, personal rejection, and emphatic understanding \citep{fitzpatrick1979you}. Further research on \textit{damage control} can evaluate the adoption of human-human strategies in human-chatbot communication.

\subsubsection{Thoroughness}
\label{sec:thoroughness}

\textit{Thoroughness} is the ability of a chatbot to be precise regarding how it uses language \citep{morrissey2013realness}. In traditional user interfaces, user communication takes place using visual affordances, such as buttons, menus, or links. In a conversational interface, language is the main tool to achieve the communicative goal. Thus, chatbots should coherently use language that portrays the expected style \citep{mairesse2009can}. When the chatbot uses inconsistent language, or unexpected patterns of language (e.g., excessive formality), the conversation may sound strange to the user, leading to frustration. 

We found 13 papers that report the importance of \textit{thoroughness} in chatbot design, three of which investigate how patterns of language influence users' perceptions and behavior toward the chatbots \citep{duijst2017can, mairesse2009can, hill2015real}. \citet{gnewuch2017towards} and \citet{morris2002conversational} suggest design principles that include concerns about language choices. Log of conversations revealed issues regarding \textit{thoroughness} in two studies \citep{jenkins2007analysis, coniam2008evaluating}. In the remaining papers, \textit{thoroughness} emerged from interviews and users' subjective feedback \citep{zamora2017sorry, morrissey2013realness, kirakowski2009establishing, tallyn2018ethnobot, chaves2018single, thies2017how}. \textit{Thoroughness} is mainly reported for open domain (five studies) and customer service chatbots (two studies), where the interactions are expected to be natural and credible to succeed \citep{morrissey2013realness, gnewuch2017towards}. \textit{Thoroughness} was also reported in another six domains of studies, such as financial services and education (see the supplementary materials).

We found two benefits of providing \textit{thoroughness}:

\textbf{[B1] to increase human-likeness:} chatbot utterances are often pre-recorded by the chatbot designer \citep{mairesse2009can}. On the one hand, this approach produces high quality utterances; on the other hand, it reduces flexibility, since the chatbot is not able to adapt the tone of the conversation based on individual users and conversational context. When analyzing interactions with a customer representative chatbot, \citet{jenkins2007analysis} observed that the chatbot proposed synonyms to keywords, and the repetition of this vocabulary led the users to imitate it. \citet{hill2015real} observed a similar tendency to matching language style. The authors compared human-human conversations with human-chatbots conversations regarding language use. They found that people use, indeed, fewer words per message and a more limited vocabulary with chatbots. However, a deeper investigation revealed that the human interlocutors were actually matching the patterns of language use with the chatbot, who sent fewer words per message. When interacting with a chatbot that uses many emojis and letter reduplication \citep{thies2017how}, participants reported a draining experience, since the chatbot's energy was too high to match. These outcomes show that adapting the language to the interlocutor is a common behavior for humans, and so chatbots would benefit from manifesting it. In addition to the interlocutor, chatbots should adapt their language use to the context in which they are implemented and adopt appropriate linguistic register \citep{morrissey2013realness, gnewuch2017towards}. In the customer services domain, \citet{gnewuch2017towards} state that chatbots are expected to fulfill the role of a human; hence, they should produce language that corresponds to the represented service provider. In the financial scenario \citep{duijst2017can}, some participants complained about the use of emojis in a situation of urgency (blocking a stolen credit card).

\textbf{[B2] to increase believability: }because people associate social qualities with machines \citep{reeves1996people}, chatbots are deemed sub-standard when users see them ``\textit{acting as a machine}'' \citep{jenkins2007analysis}. When analyzing chatbots' naturalness, \citet{morrissey2013realness} found that the formal grammatical and syntactical abilities of a chatbot are the biggest discriminators between good and poor chatbots (the other factors being \textit{conscientiousness}, \textit{manners}, and \textit{proactivity}). The authors highlight that chatbots should use consistent grammar and spelling. \citet{coniam2008evaluating} discusses how, even with English as Second Language (ESL) learners, basic grammar errors, such as pronoun confusion, diminish the value of the chatbot. In addition, \citet{morris2002conversational} states that believable chatbots also need to display unique characters through linguistic choices. In this sense, \citet{mairesse2009can} demonstrated that \textit{personality} can be expressed by language patterns. The authors proposed a computational framework to produce utterances to manifest a target \textit{personality}. The utterances were rated by experts in \textit{personality} evaluation and statistically compared against utterances produced by humans who manifest the target \textit{personality}. The outcomes show that a single utterance can manifest a believable \textit{personality} when using the appropriate linguistic form. Participants in \citet{jenkins2007analysis} described some interactions as ``\textit{robotic}'' if the chatbot repeated the keywords in the answers, reducing the interaction's naturalness. Similarly, in \citet{tallyn2018ethnobot}, participants complained about the ``\textit{inflexibility}'' of the pre-defined, handcrafted chatbot's responses and expressed the desire for it to talk ``\textit{more as a person.}''

Regarding the challenges, the surveyed literature shows the following:

\textbf{[C1] to decide how much to say: } in \citet{jenkins2007analysis}, some participants described the chatbot's utterances as not having enough detail, or being too generic; however, most of them appreciated finding answers in a sentence rather than in a paragraph. Similarly, \citet{gnewuch2017towards} argue that simple questions should not be too detailed, while important transactions require more information. In three studies \citep{chaves2018single, zamora2017sorry, duijst2017can}, participants complained about information overload and inefficiency caused by big blocks of texts. Balancing the granularity of information with the sentence length is a challenge to overcome.

\textbf{[C2] to be consistent:} chatbots should not combine different language styles. For example, in \citet{duijst2017can}, most users found it strange that emojis were combined with a certain level of formal contact. 
When analyzing the critical incidents about an open-domain interaction, \citet{kirakowski2009establishing} found that participants criticized chatbots when they used more formal language or unusual vocabulary, since general-purpose chatbots focus on casual interactions.

Despite the highlighted benefits, 
we did not find strategies to provide \textit{thoroughness}. \citet{morris2002conversational} proposed a rule-based architecture where the language choices consider the agent's \textit{personality}, emotional state, and beliefs about the social relationship among the interlocutors. However, they did not provide evidence of whether the proposed models produced the expected outcome. Although the literature in computational linguistics has proposed algorithms and statistical models to manipulate language style and matching (see e.g., \citet{prabhumoye2018style, zhang2017neural}), to the best of our knowledge, these strategies have not been evaluated in the context of chatbots' social interactions.

This section shows that linguistic choices influence users' perceptions of chatbots. The computer-mediated communication (CMC) field has a vast literature that shows language variation according to the media and its effect on social perceptions (see e.g. \citep{walther2007selective, baron1984computer}). Additionally, the cooperative principle \citep{grice1975logic}, particularly the maxim of quantity (the ability to give the appropriate amount of information), provides theoretical basis for the challenge of deciding how much to talk. Regarding adaptation and believability, researchers in sociolinguistic fields \citep{conrad2009register} have shown that language choices are influenced by personal style, dialect, genre, and register. For chatbots, the results presented in \citet{mairesse2009can} are promising, demonstrating that automatically generated language can manifest recognizable traits. Thus, further research on chatbot's \textit{thoroughness} could leverage CMC and linguistics theories to provide strategies that lead language to accomplish its purpose for a particular interactional context.

\subsubsection{Manners}
\label{sec:manners}

\textit{Manners} refer to the ability of a chatbot to manifest polite behavior and conversational habits \citep{morrissey2013realness}. Although individuals with different personalities and from different cultures may have different notions of what is considered polite (see e.g., \citet{watts2003politeness}), politeness can be more generally applied as rapport management \citep{brown2015politeness}, where interlocutors strive to control the harmony between people in discourse. A chatbot can manifest \textit{manners} by adopting speech acts such as greetings, apologies, and closings \citep{jain2018evaluating}; minimizing impositions \citep{tallyn2018ethnobot, toxtli2018understanding}, and making interactions more personal \citep{jain2018evaluating}. \textit{Manners} potentially reduces the feeling of annoyance and frustration that may lead the interaction to fail \citep{jain2018evaluating}.

We identified ten studies that report \textit{manners}, one of which directly investigates this characteristic \citep{wallis2005trouble}. In some studies \citep{toxtli2018understanding, chaves2018single, liao2018all, maurer2015benjamin}, \textit{manners} were observed in the analysis of conversational logs, where participants talked to the chatbot in polite, human-like ways. Users' feedback and interviews revealed users' expectations regarding chatbots politeness and personal behavior \citep{jain2018evaluating, jenkins2007analysis, morrissey2013realness, kirakowski2009establishing, kumar2010socially}. We identified studies reporting \textit{manners} in nine different domains, with only open domain appearing twice. The list includes education, information search, and task management, among others. See the supplementary materials for the complete list of domains of studies that report \textit{manners} as a social characteristic for chatbots.

The main benefit of providing \textit{manners} is \textbf{[B1] to increase human-likeness}. \textit{Manners} is highlighted in the literature as a way to generate a more natural, convincing interaction in chatbot conversations \citep{kirakowski2009establishing, morrissey2013realness}. In an in-the-wild data collection, \citet{toxtli2018understanding} observed that 93\% of the participants used polite words (e.g., ``\textit{thanks}'' or ``\textit{please}'') with a task management chatbot at least once, and 20\% always spoke politely to the chatbot. Unfortunately, the chatbot evaluated in that study was not prepared to handle these protocols and ultimately failed to understand. When identifying incidents from their own conversational logs with a chatbot \citep{kirakowski2009establishing}, several participants identified greetings as a human-seeming characteristic. The users also found convincing when the chatbot appropriately reacts to social cues statements, such as \textit{``how are you?''}-types of utterances. Using this result, \citet{morrissey2013realness} later suggested that greetings, apologies, social niceties, and introductions are significant constructs to measure chatbots' naturalness. In \citet{jenkins2007analysis}, the chatbot used exclamation marks at some point and frequently offered sentences available on the website, in a vaguely human-like matter. In the feedback, participants described the chatbot as rude, impolite, and cheeky. 

The surveyed literature highlights two challenges to convey \textit{manners}:

\textbf{[C1] to deal with face-threatening acts:} Face-Threatening Acts (FTA) are speech acts that threaten, either positive or negatively, the ``face'' of an interlocutor \citep{brown1987politeness}. Politeness strategies in human-human interactions are adopted to counteract the threat when an FTA needs to be performed \citep{brown1987politeness}. In \citet{wallis2005trouble}, the authors discuss that the wizard performing the role of the chatbot used several politeness strategies to counteract face threats. For instance, when she did not recognize a destination, instead of providing a list of possible destinations, she stimulated the user to keep talking until they volunteered the information. In chatbot design, by contrast, providing a list of options to choose is a common strategy. For example, in \citet{toxtli2018understanding}, the chatbot was designed to present the user with a list of pending tasks when it did not know what task the user was reporting as completed, although the authors acknowledged that it resulted in an unnatural interaction. Although adopting politeness strategies is natural for humans and people usually do not consciously think about them, implementing them for chatbots is challenging due to the complexity of identifying face-threatening acts. For example, in the decision-making coach scenario, \citet{maurer2015benjamin} observed that users tend to utter straightforward and direct agreements while most of the disagreements contained modifiers that weakened their disagreement. The adoption of politeness strategies to deal with face-threatening acts is still under-investigated in the chatbot literature.

\textbf{[C2] to end a conversation gracefully:} \citet{jain2018evaluating} discuss that first-time users expected human-like conversational etiquette from the chatbots, specifically introductory phrases and concluding phrases. Although several chatbots perform well in the introduction, the concluding phrases are less explored. Most of the participants reported being annoyed with chatbots that do not end a conversation \citep{jain2018evaluating}. \citet{chaves2018single} also highlight that chatbots need to know when the conversation ends. In that scenario, the chatbot could recognize a closing statement (the user explicitly says \textit{``thank you''} or \textit{``bye''}); however, it would not end the conversation otherwise. Users that stated a decision, but kept receiving more information from the chatbot, reported feeling confused and undecided afterward. Thus, recognizing the right moment to end the conversation is a challenge to overcome.

The strategies highlighted in the surveyed literature for providing \textit{manners} are the following:

\textbf{[S1] to engage in small talk:} \citet{liao2018all} and \citet{kumar2010socially} point out that even task-oriented chatbots engage in small talk. When categorizing the utterances from the conversational log, the authors found a significant number of messages about the agent status (e.g., \textit{``what are you doing?}''), opening and closing sentences as well as acknowledgment statements (\textit{``ok,''} \textit{``got it''}). \citet{jain2018evaluating} also observed that first-time users included small talk in the introductory phrases. According to \citet{liao2018all}, these are common behaviors in human-human chat interface, and chatbots would likely benefit from anticipating these habitual behaviors and reproducing them. However, particularly for task-oriented chatbots, it is important to control the small talk to avoid off-topic conversations and harassment, as discussed in Sections \ref{sec:damagecontrol} and \ref{sec:conscientiousness}.

\textbf{[S2] to adhere to turn-taking protocols:} \citet{toxtli2018understanding} suggest that chatbots should adopt turn-taking protocols to know when to talk. Participants who received frequent follow-up questions from the task management chatbot about their pending tasks perceived the chatbot as invasive. Literature in chatbot development proposes techniques to improve chatbots' turn-taking capabilities (see e.g., \citet{brown1987politeness, deBayser2017hybrid, candello2018having}), which can be explored as a means of improving chatbots' perceived \textit{manners}.

Although the literature emphasizes that \textit{manners} are important to approximate chatbot interactions to human conversational protocols, this social characteristic is under-investigated in the literature. Conversational acts such as greetings and apologies are often adopted (e.g., \citet{jain2018evaluating, jenkins2007analysis, maurer2015benjamin}), but there is a lack of studies on the rationality around the strategies with politeness models used in human-human social interactions \citep{wallis2005trouble}. In addition, the literature points to needs for personal conversations (e.g., addressing the users by name), but we did not find studies that focus on this type of strategy. CMC is by itself more impersonal than face-to-face conversation \citep{walther1996computer, walther1992interpersonal}; even so, current online communication media has been successfully used to initiate, develop, and maintain interpersonal relationships \citep{walther2011theories}. Researchers can learn from human behaviors in CMC and adopt similar strategies to produce more personal conversations.

\subsubsection{Moral agency}
\label{sec:moralagency}

Machine \textit{moral agency} refers to the ability of a technology to act based on social notions of right and wrong \citep{banks2018perceived}. The lack of this ability may lead to cases such as Tay, Microsoft's Twitter chatbot that became racist, sexist, and harassing in a few hours \citep{neff2016automation}. The case raised concerns on what makes an artificial agent (im)moral. Whether machines can be considered (moral) agents is widely discussed in the literature (see e.g., \citet{himma2009artificial, allen2006machine, parthemore2013makes}). In this survey, the goal is not to argue about criteria to define a chatbot as moral, but to discuss the benefits of manifesting a perceived agency \citep{banks2018perceived} and the implications of disregarding chatbots' moral behavior. Hence, for the purpose of this survey, \textit{moral agency} is a manifested behavior that may be inferred by a human as morality and agency \citep{banks2018perceived}.

We found six papers that address \textit{moral agency}. \citet{banks2018perceived} developed and validated a metric for perceived \textit{moral agency} in conversational interfaces, including chatbots. In four studies, the authors investigated the ability of chatbots to handle conversations where the persistence of gender \citep{deangeli2006sex, brahnam2012gender} and racial stereotypes \citep{marino2014racial, schlesinger2018let} may occur. In \citet{shum2018eliza}, \textit{moral agency} is discussed as a secondary result, where the authors discuss the impact of biased responses on emotional connection. \textit{Moral agency} was observed in only two domains of studies: open domain (four studies) and race-talk (two studies), which shows that this characteristic is primarily relevant when the conversational topic may raise moral concerns, which ultimately requires ethical behavior from the conversational partners.

The two main reported benefits of manifesting perceived \textit{moral agency} are the following:

\textbf{[B1] to avoid stereotyping:} chatbots are often designed with anthropomorphized characteristics (see Section \ref{sec:personification}), including gender, age, and ethnicity identities. Although the chatbot's personification is more evident in embodied conversational agents, text-based chatbots may also be assessed by their social representation, which risks building or reinforcing stereotypes \citep{marino2014racial}. \citet{marino2014racial} and \citet{schlesinger2018let} argue that chatbots are often developed using language registers \citep{marino2014racial} and cultural references \citep{schlesinger2018let} of the dominant culture. In addition, a static image (or avatar) representing the agent may convey social grouping \citep{nowak2005influence}. When the chatbot is positioned in a minority \textit{identity} group, it exposes the image of that group to judgment and flaming, which is frequent in chatbot interactions \citep{marino2014racial}. For example, \citet{marino2014racial} discusses the controversies caused by a chatbot designed to answer questions about Caribbean Aboriginal culture: its representation as a Caribbean Amerindian individual created an unintended context for stereotyping, where users projected the chatbot's behavior as a standard for people from the represented population. Another example is the differences in sexual discourse between male- and female-presenting chatbots. \citet{brahnam2012gender} found that female-presenting chatbots are the object of implicit and explicit sexual attention and swear words more often than male-presenting chatbots. \citet{deangeli2006sex} show that sex talks with the male chatbot were rarely coercive or violent; his sexual preference was often questioned, though, and he was frequently propositioned by reported male users. In contrast, the female character received violent sexual statements, and was threatened with rape five times in the analyzed corpora. In \citet{brahnam2012gender}, when the avatars were presented as black adults, references to race often deteriorated into racist attacks. Manifesting moral agency may, thus, prevent obnoxious user interactions. In addition, moral agency may prevent the chatbot itself from being biased or disrespectful to humans. \citet{schlesinger2018let} argue that the lack of context about the world does not redeem the chatbot from the necessity of being respectful with all the social groups.

\textbf{[B2] to enrich interpersonal relationships:} in a study on how interlocutors perceive conversational agents' \textit{moral agency}, \citet{banks2018perceived} hypothesized that perceived morality may influence a range of motivations, dynamics, and effects of human-machine interactions. Based on this claim, the authors evaluated whether goodwill, trustworthiness, willingness to engage, and relational certainty in future interactions are constructs to measure perceived \textit{moral agency}. Statistical results showed that all the constructs correlate with morality, which suggests that manifesting \textit{moral agency} can enrich interpersonal relationships with chatbots. Similarly, \citet{shum2018eliza} suggest that to produce interpersonal responses, chatbots should be aware of inappropriate information and avoid generating biased responses.

However, the surveyed literature also reveals challenges of manifesting \textit{moral agency}:

\textbf{[C1] to avoid alienation:} in order to prevent a chatbot from reproducing hate speech or abusive talk, most chatbots are built over ``clean'' data, where specific words are removed from their dictionary \citep{schlesinger2018let, deangeli2006sex}. These chatbots have no knowledge of those words and their meaning. Although this strategy is useful to prevent unwanted behavior, it does not manifest agency, but rather alienates the chatbot of the topic. \citet{deangeli2006sex} show that the lack of understanding about sex-talk does not prevent the studied chatbot from harsh verbal abuse, or even from being perceived as encouraging such abuse. From \citep{schlesinger2018let}, one can notice that the absence of racist specific words did not prevent the chatbot Zo from uttering discriminatory exchanges. As a consequence, manifesting \textit{moral agency} requires a broader understanding of the world, rather than alienation, which is an open challenge.

\textbf{[C2] to build unbiased algorithms and training data:} as extensively discussed in \citet{schlesinger2018let}, machine learning algorithms and corpus-based language generation are biased toward the available training datasets. Hence, \textit{moral agency} relies on data that is biased in its nature, producing unsatisfactory results from an ethical perspective. In \citet{shum2018eliza}, the authors propose a framework for developing social chatbots. The authors highlight that the core chat module should follow ethical design to generate unbiased, non-discriminative responses, but they do not discuss specific strategies for that. Building unbiased training datasets and learning algorithms that connect the outputs with individual, real-world experiences, therefore, are challenges to overcome.

Despite the relevance of \textit{moral agency} to the development of socially intelligent chatbots, we did not find strategies to address the issues. \citet{schlesinger2018let} advocate for developing diversity-conscious databases and learning algorithms that account for ethical concerns; however, the paper focuses on outlining the main research branches and calls on the community of designers to adopt new strategies. As discussed in this section, research on perceived \textit{moral agency} is still necessary in order to develop chatbots whose social behavior is inclusive and respectful. In the field of embodied agents, mind perception theory \citep{gray2007dimensions} has been investigated as a means to improve interactions through agency and emotion \citep{lee2019virtual, appel2020uncanny, keijsers2018mindless}. Future investigations in chatbots interactions could leverage this theory to understand when and the extent to which perceived moral agency improves communication with chatbots.

\subsubsection{Emotional Intelligence}
\label{sec:emotionalintelligence}

\textit{Emotional intelligence} is a subset of social intelligence that allows an individual to appraise and express feelings, regulate affective reactions, and harness emotions to solve a problem \citep{salovey1990emotional}. Although chatbots do not have genuine emotions \citep{wallis2005trouble}, there are considerable discussions about the role of manifesting emotional cues in chatbots \citep{shum2018eliza, wallis2005trouble, ho2018psychological}. An emotionally intelligent chatbot can recognize and influence users' feelings and demonstrate respect, empathy, and understanding, improving the human-chatbot relationship \citep{salovey1990emotional, li2017dailydialog}.

We identified 14 studies that report \textit{emotional intelligence}. Unlikely the previously discussed categories, most studies on \textit{emotional intelligence} focused on understanding the effects of chatbots' empathy and emotional self-disclosure \citep{ayedoun2017communication, kumar2010socially, fitzpatrick2017delivering, dohsaka2014effects, ho2018psychological, shum2018eliza, miner2016conversational, morris2002conversational, portela2017new, lee2017enhancing, wallis2005trouble}. Only three papers highlighted \textit{emotional intelligence} as an exploratory outcome \citep{thies2017how, jenkins2007analysis, zamora2017sorry}, where needs for \textit{emotional intelligence} emerged from participants' subjective feedback and post-intervention surveys. \textit{Emotional intelligence} is mainly investigated in domains where topics may involve the disclosure of feelings (e.g., in open domain interactions \citep{shum2018eliza}) and expressions of empathy and understanding are appropriate \citep{fitzpatrick2017delivering, dohsaka2014effects} (e.g., health care, gaming, education). See the supplementary materials for the domains of study investigating \textit{emotional intelligence}.

The main reported benefits of developing emotionally intelligent chatbots are the following:

\textbf{[B1] to enrich interpersonal relationships:} the perception that the chatbot understands one's feelings may create a sense of belonging and acceptance \citep{ho2018psychological}. \citet{ayedoun2017communication} propose that chatbots for second language studies should use congratulatory, encouraging, sympathetic, and reassuring utterances to create a friendly atmosphere to the learner. The authors statistically demonstrated that affective backchannel, combined with communicative strategies (see Section \ref{sec:conscientiousness}) significantly increased learners' confidence and desire to communicate, while reducing anxiety. In another educational study, \citet{kumar2010socially} evaluated the impact of chatbot's affective moves on friendliness and achieving social belonging. Qualitative results show that affective moves significantly improve the perception of amicability, and marginally increased social belonging. According to \citet{wallis2005trouble}, when a chatbot's emotional reaction triggers a social response from the user, the chatbot has achieved group membership and users' sympathy. \citet{dohsaka2014effects} proposed a chatbot that uses empathic and self-oriented emotional expressions to keep users engaged in quiz-style dialogue. The survey results revealed that empathic expressions significantly improved user satisfaction. In addition, the empathic expressions also improved the user ratings of the peer agent regarding intimacy, compassion, amiability, and encouragement. Although \citet{dohsaka2014effects} did not find effect of chatbot's self-disclosure on emotional connection, \citet{lee2017enhancing} found that self-disclosure and reciprocity significantly improved trust and interactional enjoyment. In \citet{fitzpatrick2017delivering}, seven participants reported that the best thing about their experience with the therapist chatbot was perceived empathy. Five participants highlighted that the chatbot demonstrated attention to their users' feelings. In addition, the users referred the chatbot as \textit{``he,'' ``a friend,''} \textit{``a fun little dude,''} which demonstrates that empathy emerged from the personification of the chatbot. In another mental health care study, \citet{miner2016conversational} found that humans are twice as likely to mirror negative sentiment from a chatbot than from a human, which is a relevant implication for therapeutic interactions. In \citet{zamora2017sorry}, participants reported that some content is embarrassing to ask another human, thus, talking to a chatbot would be easier due to the lack of judgement. \citet{ho2018psychological} measured users' experience in conversations with a chatbot compared to a human partner as well as the amount of intimacy disclosure and cognitive reappraisal. Participants in the chatbots condition experienced as many emotional, relational, and psychological benefits as participants who disclosed to a human partner. 

\textbf{[B2] to increase engagement:} \citet{shum2018eliza} argue that longer conversations (10+ turns) are needed to fulfill the needs of affection and belonging. Therefore, the authors defined conversation-turns per session as a success metric for chatbots, where usefulness and emotional understanding are combined. In \citet{dohsaka2014effects}, empathic utterances for the quiz-style interaction significantly increase the number of users' messages per hint for both answers and non-answer utterances (such as feedback about the success/failure). This result shows that empathic utterances encouraged the users to engage and utter non-answers statements. \citet{portela2017new} compared the possibility of emotional connection between a classical chatbot and a pretend chatbot, simulated in a WoZ experiment. Quantitative results showed that the WoZ condition was more engaging, since it resulted in conversations that lasted longer, with a higher number of turns. The analysis of the conversational logs revealed the positive effect of the chatbot manifesting social cues and empathic signs as well as touching on personal topics.

\textbf{[B3] to increase believability:} \citep{morris2002conversational} argue that adapting chatbots' language to their current emotional state, along with their \textit{personality} and social role awareness, results in more believable interactions. The authors propose that conversation acts should reflect the pretended emotional status of the agent; the extent to which the acts impact on emotion depends however on the agent's \textit{personality} (e.g., its temperament or tolerance). \textit{Personality} is an anthropomorphic characteristic and is discussed in Section \ref{sec:personality}.

Although \textit{emotional intelligence} is the goal of several studies, \textbf{[C1] regulating affective reactions} is still a challenge. The chatbot presented in \citet{kumar2010socially} was designed to mimic the patterns of affective moves in human-human interactions. Nevertheless, the chatbot has shown an only marginally significant increase in social belonging, when compared to the same interaction with a human partner. Conversational logs revealed that the human tutor performed a significantly higher number of affective moves in that context. In \citet{jenkins2007analysis}, the chatbot was designed to present emotive-like cues, such as exclamation marks, and interjections. The participants negatively rated the degree of emotion in the chatbot's responses. In \citet{thies2017how}, the energetic chatbot was reported as having an enthusiasm too high to match. In contrast, the chatbot was described as an \textit{``emotional buddy''} was reported as being \textit{``overly caring.'' } \citet{ho2018psychological} state that chatbot's empathic utterances may be seen as pre-programmed and inauthentic. Although their results revealed that the partners' identity (chatbot vs. person) had no effect in the perceived relational and emotional experience, the chatbot condition was a WoZ setup. The wizards were blind to whether users thought they were talking to a chatbot or a person, which reveals that identity does not matter if the challenge of regulating emotions is overcome.

The chatbots literature also report some strategies to manifest \textit{emotional intelligence}:

\textbf{[S1] using social-emotional utterances:} affective utterances toward the user are a common strategy to demonstrate \textit{emotional intelligence}. \citet{ayedoun2017communication}, \citet{kumar2010socially}, and \citet{dohsaka2014effects} suggest that affective utterances improve the interpersonal relationship with a tutor chatbot. In \citet{ayedoun2017communication}, the authors propose affective backchannel utterances (congratulatory, encouraging, sympathetic, and reassuring) to motivate the user to communicate in a second language. The tutor chatbot proposed in \citet{kumar2010socially} uses solidarity, tension release, and agreement utterances to promote its social belonging and acceptance in group chats. \citet{dohsaka2014effects} propose empathic utterances to express opinion about the difficulty or ease of a quiz, and feedback on success and failure.

\textbf{[S2] to manifest \textit{conscientiousness}:} demonstrating \textit{conscientiousness} may affect the emotional connection between humans and chatbots. In \citet{portela2017new}, participants reported the rise of affection when the chatbot remembered something they had said before, even if it was just the user's name. Keeping track of the conversation was reported as an empathic behavior and resulted in mutual affection. \citet{shum2018eliza} argue that a chatbot needs to combine usefulness with emotion by asking questions that help to clarify the users' intentions. They provide an example where a user asks the time, and the chatbot answer \textit{``Cannot sleep?''} as an attempt to guide the conversation to a more engaging direction. Adopting this strategy requires the chatbot to handle message understanding, emotion and sentiment tracking, session context modeling, and user profiling \citet{shum2018eliza}.

\textbf{[S3] reciprocity and self-disclosure:} \citet{lee2017enhancing} hypothesized that a high level of self-disclosure and reciprocity in communication with chatbots would increase trust, intimacy, and enjoyment, ultimately improving user satisfaction and intention to use. They performed a WoZ, where the assumed chatbot was designed to recommend movies. Results demonstrated that reciprocity and self-disclosure are strong predictors of rapport and user satisfaction. In contrast, \citet{dohsaka2014effects} did not find any effect of self-oriented emotional expressions in the users' satisfaction or engagement. More research is needed to understand the extent to which this strategy produces positive impact on the interaction.

The literature shows that \textit{emotional intelligence} is widely investigated, with particular interest from education and mental health care domains. Using emotional utterances in a personalized, context relevant way is still a challenge. Researchers in chatbots \textit{emotional intelligence} can learn from \textit{emotional intelligence} theory \citep{salovey1990emotional, gross1998emerging} to adapt the chatbots utterances to match the emotions expressed in the dynamic context. Adaption to the dynamic context also improves the sense of personalized interactions, which is discussed in the next section.

\subsubsection{Personalization}
\label{sec:personalization}

\textit{Personalization} refers to the ability of a technology to adapt its functionality, interface, information access, and content to increase its personal relevance to an individual or a category of individuals \citep{fan2006personalization}. In the chatbot domain, \textit{personalization} may increase the agents' social intelligence, since it allows a chatbot to be aware of situational context and dynamically adapt its features to better suit individual needs \citep{neururer2018perceptions}. Grounded in robots and artificial agents' literature, \citet{liao2016what} argue that \textit{personalization} can improve rapport and cooperation, ultimately increasing engagement with chatbots. Although some studies (see e.g., \citet{zhang2018personalizing, fan2006personalization, liao2016what}) also relate \textit{personalization} to the attribution of personal qualities such as \textit{personality}, we discuss personal qualities in the \textbf{Personification} category. In this section, we focus on the ability to adapt the interface, content, and behavior to the users' preferences, needs, and situational context.

We found 11 studies that report \textit{personalization}. Three studies pose \textit{personalization} as a research goal \citep{duijst2017can, liao2016what, shum2018eliza}. In most of the studies, though, \textit{personalization} was observed in exploratory findings. In six studies, \textit{personalization} emerged from the analysis of interviews and participants' self-reported feedback \citep{neururer2018perceptions, duijvelshoff2017use, tallyn2018ethnobot, thies2017how, portela2017new, jenkins2007analysis}. In two studies \citep{lasek2013chatbots, toxtli2018understanding}, needs for \textit{personalization} emerged from the conversational logs. \textit{Personalization} was investigated in seven different domains, including open domain and task management (two studies each). In open domain interactions, personalization is derived from remembering information from previous interactions, such as personal preferences and users' details \citep{thies2017how}. In task-oriented contexts, such as task management, personalization aims to increase the relevance of services to particular users \citep{liao2016what}. See the supplementary materials for details.

The surveyed literature highlighted three benefits of providing personalized interactions:

\textbf{[B1] to enrich interpersonal relationships:} \citet{duijvelshoff2017use} state that personalizing the amount of personal information a chatbot can access and store is required to establish a relation of trust and reciprocity in workplace environments. In \citet{neururer2018perceptions}, interviews with 12 participants generated a total of 59 statements about how learning from experience promotes chatbots' authenticity. \citet{shum2018eliza} argue that chatbots whose focus is engagement need to personalize the generation of responses for different users' backgrounds, personal interests, and needs in order to serve their needs for communication, affection, and social belonging. In \citet{portela2017new}, participants expressed the desire for the chatbot to provide different answers to different users. Although \citet{duijst2017can} has found no significant effect of \textit{personalization} on the user experience with the financial assistant chatbot, the study applies \textit{personalization} as the ability to provide empathic responses according to the users' issues, where \textit{emotional intelligence} plays a role. Interpersonal relationship can also be enriched by adapting the chatbots' language to match the user's context, energy, and formality; the ability to appropriately use language is discussed in Section \ref{sec:thoroughness}.

\textbf{[B2] to provide unique services:} providing \textit{personalization} increases the value of provided information \citep{duijst2017can}. In the ethnography data collection study \citep{tallyn2018ethnobot}, eight participants reported dissatisfaction with the chatbot's generic guidance to specific places. Participants self-reported that the chatbot should use their current location to direct them to places more conveniently located, and ask for participants interests and preferences to direct them to areas that meet their needs. When exploring how teammates used a task-assignment chatbot, \citet{toxtli2018understanding} found that the use of the chatbot varied depending on the participants' levels of hierarchy. Similarly, qualitative analysis of perceived interruption in a workplace chat \citep{liao2016what} suggests that interruption is likely associated with users' general aversion to unsolicited messages at work. Hence, the authors argue that chatbot's messages should be personalized to the user's general preference. \citet{liao2016what} also found that users with low social-agent orientation emphasize the utilitarian value of the system, while users with high social-agent orientation see the system as a humanized assistant. This outcome support to the need to personalize the interaction to individual user's mental models. In \citet{thies2017how}, participants reported preference for a chatbot that remembers their details, likes and dislikes, and preferences, and voluntarily uses the information to make recommendations. In \citet{jain2018evaluating}, two participants also expected chatbots to retain context from previous interactions to improve recommendations.

\textbf{[B3] to reduce interactional breakdowns:} in HCI, \textit{personalization} is used to customize the interface toward user familiarity \citep{fan2006personalization}. When evaluating
visual elements (such as quick replies) compared to typing the responses, \citet{lasek2013chatbots} observed that users that start the interaction by clicking an option are more likely to continue the conversation if the next exchange also has visual elements as optional affordances. In contrast, users who typed are more likely to abandon the conversation when they face options to click. Thus, chatbots should adapt their interface to users' preferred input methods. In \citet{jenkins2007analysis}, one participant suggested that the choice of text color and font size should be customizable. \citet{duijst2017can} also observed that participants faced difficulties with small letters, and concluded that adapting the interface to provide accessibility also needs to be considered.

According to the surveyed literature, the main challenge regarding \textit{personalization} is \textbf{[C1] privacy}. To enrich the efficiency and productivity of the interaction, a chatbot needs to have memory of previous interactions as well as learn user's preferences and disclosed personal information \citep{thies2017how}. However, as \citet{duijvelshoff2017use} and \citet{thies2017how} suggest, collecting personal data may lead to privacy concerns. Thus, chatbots should showcase transparent purpose and ethical standards \citep{neururer2018perceptions}. \citet{thies2017how} also suggest that there should be a way to inform a chatbot that something in the conversation is private. Similarly, participants in \citet{zamora2017sorry} reported that personal data 
and social media content may be inappropriate topics for chatbots because they can be sensitive
. These concerns may be reduced if a chatbot demonstrates care about privacy \citep{duijvelshoff2017use}.

The reported strategies to provide \textit{personalization} in chatbots interactions are the following:

\textbf{[S1] to learn from and about the user:} \citet{neururer2018perceptions} state that chatbots should present strategies to learn from cultural, behavioral, personal, conversational, and contextual interaction data. For example, the authors suggest using Facebook profile information to build knowledge about users' personal information. \citet{thies2017how} also suggest that the chatbot should remember user's preferences disclosed in previous conversations. In \citet{shum2018eliza}, the authors propose an architecture where responses are generated based on a \textit{personalization} rank that applies users' feedback about their general interests and preferences. When evaluating the user's experience with a virtual assistant chatbot, \citet{zamora2017sorry} found 16 mentions to personalized interactions, where participants demonstrated the need for a chatbot to be aware of their personal quirks and anticipate their needs.

\textbf{[S2] to provide customizable agents:} \citet{liao2016what} suggest that users should be able to choose the level of the chatbot's attributes, for example, the agent's look and persona. By doing so, users with low social-agent orientation could use a non-humanized interface, which would better represent their initial perspective. This differentiation could be the first signal to personalize further conversation, such as focusing on more productive or playful interactions. Regarding chatbots' learning capabilities, in \citet{duijvelshoff2017use}, interviews with potential users revealed that users should be able to manage what information the chatbot knows about them and decide whether the chatbot can learn from previous interactions or not. If the user prefers a more generic chatbot, then it would not store personal data, potentially increasing the engagement with more resistant users. \citet{thies2017how} raise the possibility of offering an ``incognito'' mode for chatbots or asking the chatbot to forget what was said in previous utterances.

\textbf{[S3] visual elements:} \citet{tallyn2018ethnobot} adopted quick replies as a means for the chatbot to tailor its subsequent questions to the specific experience the participant had reported. As discussed in Section \ref{sec:conscientiousness}, quick replies may be seen as restrictive from an interactional perspective; however, conversation logs showed that the tailored questions prompted the users to report more detail about their experience, which is important in ethnography research. 

Both the benefits and strategies identified in the literature are in line with the types of \textit{personalization} proposed by \citet{fan2006personalization}. Therefore, further investigations in \textit{personalization} can leverage knowledge from interactive systems (e.g., \citet{fan2006personalization, thomson2005standard}) to adapt \textit{personalization} strategies and manage the privacy concern.

\begin{framed} \small 
\vspace{-2mm}
In summary, the \textbf{social intelligence} category includes characteristics that help a chatbot to manifest an adequate social behavior, by managing conflicts, using appropriate language, displaying \textit{manners} and \textit{moral agency}, sharing emotions, and handling personalized interactions. The benefits relate to resolving social positioning and recovering from failures, as well as increasing believability, human-likeness, engagement, and rapport. To achieve that, designers and researchers should care about privacy, emotional regulation issues, language consistency, and identification of failures and inappropriate content.
\vspace{-2mm}
\end{framed}

\subsection{Personification}
\label{sec:personification}

In this section, we discuss the influence of \textit{identity} projection on human-chatbot interaction. \textbf{Personification} refers to assigning personal traits to non-human agents, including physical appearance, and emotional states \citep{fan2006personalization}. In the HCI field, researchers argue that using a personified character in the user interface is a natural way to support the interaction \citep{koda2003user}. Indeed, the literature shows that (i) users can be induced to behave as if computers were humans, even when they consciously know better \citep{nass1993anthropomorphism}; and (ii) the more human-like a computer representation is, the more social people's responses will be \citep{gong2008social}.

Chatbots are, by definition, designed to have at least one human-like trait: (human) natural language. Although research on \textbf{personification} is more common in the Embodied Conversational Agents field, \citet{deangeli2001unfriendly} claim that chatbot embodiment can be created through narrative without any visual help. According to \citet{deangeli2005rescue}, talking to a machine affords it a new \textit{identity}. In this section, we divided the social characteristics that reflect \textbf{personification} into \textit{identity} (16 papers) and \textit{personality} (12 papers). In this category, we found several studies where part of the main investigation relates to the social characteristics. Six studies applied quantitative methods, whereas the majority reported qualitative (10) or mixed (11) methods. See the supplementary materials for details.

\subsubsection{Identity}
\label{sec:identity}

\textit{Identity} refers to the ability of an individual to demonstrate belonging to a particular social group \citep{stets2000identity}. Although chatbots lack the agency to choose what social group to which they want to belong, designers attribute \textit{identity} to them, intentionally or not, when they define the way a chatbot talks or behaves \citep{cassell2009social}. The \textit{identity} of a partner (even if only perceived) gives rise to new processes, expectations, and effects that influence the outcomes of the interaction \citep{ho2018psychological}. Aspects that convey the chatbots' \textit{identity} include gender, age, language style, and name. Additionally, chatbots may have anthropomorphic, zoomorphic, or robotic representations. Some authors include \textit{identity} aspects in the definition of \textit{personality} (see, e.g., \citet{shum2018eliza}). We distinguish these two characteristics, where \textit{identity} refers to appearance and cultural traits while \textit{personality} focuses on behavioral traits.

We found 16 studies that discuss \textit{identity} issues, ten of which include \textit{identity} as part of their main investigation \citep{liao2018all, jenkins2007analysis,  candello2017typefaces, ciechanowski2018shades, deangeli2001unfriendly, deangeli2006sex, schlesinger2018let, marino2014racial, corti2016co, araujo2018living}. In two studies, the authors argue for the impact of \textit{identity} on the interaction based on the literature \citep{gnewuch2017towards, brandtzaeg2018chatbots}. In three studies \citep{silvervarg2013iterative, thies2017how, toxtli2018understanding, neururer2018perceptions}, qualitative analysis of conversational logs revealed that participants put effort into understanding aspects of the chatbots' \textit{identity}. \textit{Identity} concerns were primarily investigated for open domain and customer service chatbots (four studies each). In open domain interactions, \textit{identity} is explored as a means of building common ground \citep{deangeli2001unfriendly}. In the case of customer services, \textit{identity} helps manifest credibility and trust \citep{gnewuch2017towards}. Other domains include race-talk, education, and gaming. The complete list appears in the supplementary materials.

The identified benefits of attributing \textit{identity} to a chatbot are the following:

\textbf{[B1] to increase engagement: }when evaluating signals of playful interactions, \citet{liao2018all} found that agent-oriented conversations (asking about an agent's traits and status) are consistent with the tendency to anthropomorphize the agent and engage in chit-chat. In the educational scenario, \citet{silvervarg2013iterative} also observed questions about agents' appearance, intellectual capacities, and sexual orientation, although the researchers considered these questions inappropriate for the context of tutoring chatbots. When comparing human-like vs. machine-like language style, greetings, and framing, \citet{araujo2018living} noticed that using informal language, having a human name, and using greetings associated with human communication resulted in significantly higher scores for adjectives like likeable, friendly, and personal. In addition, framing the agent as ``intelligent'' also had a slight influence on users' scores.

\textbf{[B2] to increase human-likeness:} some attributes may convey a perceived human-likeness. \citet{araujo2018living} showed that using human-like language style, name, and greetings resulted in significantly higher scores for naturalness. The chatbot's framing influenced the outcomes when combined with other anthropomorphic clues. When evaluating different typefaces for a financial adviser chatbot, \citet{candello2017typefaces} found that users perceive machine-like typefaces as more chatbot-like, although they did not find strong evidence of handwriting-like typefaces conveying humanness.

The surveyed literature also highlights challenges regarding \textit{identity}:

\textbf{[C1] to avoid negative stereotypes: }when engaging in a conversation, interlocutors base their behavior on common ground (joint knowledge, background facts, assumptions, and beliefs that participants have of each other) (see \citep{deangeli2001unfriendly}). Common ground reflects stereotypical attributions that chatbots should be able to manage as the conversation evolves \citep{deangeli2005rescue}. In \citet{deangeli2001unfriendly}, the authors discuss that chatbots for company representations are often personified as attractive human-like women acting as spokespeople for their companies, while men chatbots tend to have a more important position, such as a virtual CEO. \citet{deangeli2006sex} state that the agent self-disclosure of gender \textit{identity} opens possibilities to sex talk. The authors observed that the conversations mirror stereotyped male/female encounters, and the ambiguity of the chatbot's gender may influence the exploration of homosexuality. However, fewer instances were observed of sex-talk with the chatbot personified as a robot, which demonstrates that the gender \textit{identity} may lead to the stereotypical attributions. When evaluating the effect of gender identity on disinhibition, \citet{brahnam2012gender} showed that people spoke more often about physical appearance and used more swear and sexual words with the female-presenting chatbot, and racist attacks were observed in interactions with black-representing chatbots. The conversation logs from \citep{jenkins2007analysis} also show instances of users attacking the chatbot persona (a static avatar of a woman pointing to the conversation box). \citet{marino2014racial} and \citet{schlesinger2018let} also highlight that race identity conveys not only the physical appearance, but all the socio-cultural expectations about the represented group (see discussion in Section \ref{sec:moralagency}). Hence, designers should care about the impact of attributing an 
identity to chatbots in order to avoid reinforcing negative stereotypes.

\textbf{[C2] to balance the \textit{identity} and the technical capabilities:} the literature comparing embodied vs. disembodied conversational agents yields contradictory results regarding the relevance of a human representation. For example, in the context of general-purpose interactions, \citet{corti2016co} show that people exert more effort toward establishing common ground with the agent when they are represented as fully human; in contrast, when evaluating a website assistant chatbot, \citet{ciechanowski2018shades} show that simpler text-based chatbots with no visual, human \textit{identity} resulted in less of an uncanny effect and reduced negative affect. Overly humanized agents create a higher expectation for users, which eventually leads to more frustration when the chatbot fails \citep{gnewuch2017towards}. When arguing about why chatbots fail, \citet{brandtzaeg2018chatbots} advocate for balancing human versus robotic aspects, where \textit{``too human''} representations may lead to off-topic conversations, and overly robotic interactions may lack personal touch and flexibility. When arguing about the social presence conveyed by deceptive chatbots, \citet{deangeli2005rescue} state that extreme anthropomorphic features may generate cognitive dissonance. The challenge thus lies in designing a chatbot that provides appropriate \textit{identity} cues, which correspond to their capabilities and communicative purpose, in order to convey the right expectation and minimize discomfort from over-personification.

Regarding the strategies, the surveyed literature suggests \textbf{[S1] to design and elaborate on a persona}. Chatbots should have a comprehensive persona and answer agent-oriented conversations with a consistent description of itself \citep{liao2018all, neururer2018perceptions}. For example, \citet{deangeli2005rescue} discuss that Eliza, the psychotherapist chatbot, and Parry, a paranoid chatbot, have behaviors consistent with the stereotypes associated with the professional and personal identities, respectively. \citet{toxtli2018understanding} suggest that designers should explicitly build signals of the chatbot personification (either machine- or human-like), so the users can have the right expectation about the interaction. When \textit{identity} aspects are not explicit, users try to establish common ground. In \citet{liao2018all} and \citep{silvervarg2013iterative}, much of the small talk with the chatbot related to the chatbot's traits and status. In \citet{deangeli2001unfriendly}, the authors observed many instances of Alice's self-references to ``\textit{her}'' artificial nature. These references triggered the users to reflect on their human-condition (self-categorization process), resulting in exchanges about their species (either informational or confrontational). \citet{thies2017how} observed similar results, as participants engaged in conversations about the artificial nature of the agent. Providing the chatbot with the ability to describe its personal \textit{identity} helps to establish the common ground, and hence, enrich the interpersonal relationship \citep{deangeli2001unfriendly}.

Chatbots may be designed to deceive users about their actual \textit{identity}, pretending to be a human \citep{deangeli2005rescue}. In this case, the more human the chatbot sounds, the more successful it will be. In many cases, however, there is no need to engage in deception and the chatbots can be designed to represent an elaborated persona. 
Researchers can explore social identity theory \citep{stets2000identity, brown2000social} related to ingroup bias, power relations, homogeneity, and stereotyping, in order to design chatbots with \textit{identity} traits that reflect their expected social position \citep{harre2003self}.

\subsubsection{Personality}
\label{sec:personality}

\textit{Personality} refers to personal traits that help to predict someone's thoughts, feelings, and behaviors \citep{mccrae1997personality}. The most accepted set of traits is called the Five-Factor model (or Big Five model) \citep{goldberg1990alternative, mccrae1997personality}), which describes \textit{personality} across five dimensions (extraversion, agreeableness, conscientiousness, neuroticism, and openness). However, \textit{personality} can also refer to other dynamic, behavioral characteristics, such as temperament and sense of humor \citep{zuckerman1993comparison, thorson1993sense}. In the chatbots domain, \textit{personality} refers to the set of traits that determines the agent's interaction style, describes its character, and allows the end-user to understand its general behavior \citep{deangeli2001unfriendly}. Chatbots with consistent \textit{personality} are more predictable and trustworthy \citep{shum2018eliza}. According to \citet{deangeli2001personifying}, unpredictable swings in chatbots' attitudes can disorient users and create a strong sense of discomfort. Thus, \textit{personality} ensures that a chatbot displays behaviors that stand in agreement with the users' expectations in a particular context \citep{petta1997create}.

We found 12 studies that report \textit{personality} issues for chatbots. In some studies, \textit{personality} was investigated in reference to the Big Five model \citep{morris2002conversational, sjoden2011extending, mairesse2009can}, while two studies focused on sense of humor \citep{ptaszynski2010towards, meany2010humour}. Three studies investigated the impact of the \textit{personality} of tutor chatbots on students' engagement \citep{sjoden2011extending, ayedoun2017communication, kumar2010socially}. \citet{thies2017how} compared users' preferences regarding pre-defined personalities. In the remaining studies, \citep{portela2017new, shum2018eliza, brandtzaeg2017people, jain2018evaluating} \textit{personality} concerns emerged from the qualitative analysis of the interviews, users' subjective feedback, and literature reviews \citep{meany2010humour}. \textit{Personality} was mostly investigated in open domain (five studies) and education (three studies) chatbots. The other two domains were gaming and humorous chatbots, which are both playful agents. Hence, \textit{personality} is relevant when believability and attitude play a role in the interaction \citep{portela2017new} (e.g., in open domain) and when the chatbots' attitude may increase users' mental comfort when performing a task \citep{ayedoun2017communication}, such as in educational contexts.

The surveyed literature revealed two benefits of attributing \textit{personality} to chatbots:

\textbf{[B1] to increase believability:} \citet{morris2002conversational} states that chatbots should have a \textit{personality}, defined by the Five Factor model, plus characteristics such as temperament and tolerance, in order to build utterances using linguistic choices that cohere with these attributions. When evaluating a humorous chatbot, \citet{ptaszynski2010towards} compared the naturalness of the chatbot's inputs and the chatbot's human-likeness compared to a non-humorous chatbot. The humorous chatbot scored significantly higher in both constructs. \citet{portela2017new} also showed that sense of humor humanizes the interactions, since humor was one of the factors that influenced naturalness for the WoZ condition. \citet{mairesse2009can} demonstrated that manipulating language to manifest a target \textit{personality} produced moderately natural utterances, with a mean rating of 4.59 out of 7 for the \textit{personality} model utterances.

\textbf{[B2] to enrich interpersonal relationships:} chatbots \textit{personality} can make the interaction more enjoyable \citep{brandtzaeg2017people, jain2018evaluating}. In \citet{brandtzaeg2017people}, the second most frequent motivation for using chatbots, noted by 20\% of the participants, was entertainment. The authors argue that a chatbot's capacity to be fun is important even when the main purpose is productivity; according to participants, the chatbot's ``\textit{fun tips}'' enrich the user experience. This result is consistent with the experience of first-time users \citep{jain2018evaluating}, where participants relate better with chatbots who have consistent \textit{personality}. \citet{thies2017how} show how witty banter and casual, enthusiastic conversations help make the interaction effortless. In addition, a few participants enjoyed the chatbot with a caring \textit{personality}, who was described as a good listener. In \citet{shum2018eliza} and \citep{sjoden2011extending}, the authors argue that a consistent \textit{personality} helps the chatbot to gain the users' confidence and trust. \citet{sjoden2011extending} state that tutor chatbots should display appropriate posture, conduct, and representation, which include being encouraging, expressive, and polite. Accordingly, other studies highlight students' desire for chatbots with positive agreeableness and extraversion \citep{sjoden2011extending, ayedoun2017communication, kumar2010socially}. Outcomes consistently suggest that students prefer a chatbot that is not overly polite, but has some attitude. Agreeableness seems to play a critical role, helping the students to feel encouraged and deal with difficulties. Notably, agreeableness requires \textit{emotional intelligence} to be warm and sympathetic in appropriate circumstances (see Section \ref{sec:emotionalintelligence}).

The reviewed literature pointed out two challenges regarding \textit{personality}:

\textbf{[C1] to adapt humor to the users' culture:} sense of humor is highly shaped by cultural environment \citep{ruch1998sense}. \citet{ptaszynski2010towards} discusses a Japanese chatbot who uses puns to create funny conversations. The authors state that puns are one of the main humor genres in that culture. However, puns are restricted to a specific culture and language, with low portability. Thus, the design challenge lies in personalizing chatbots' sense of humor to the target users' culture and interests or, alternatively, designing cross-cultural forms of humor. The ability to adapt to the context and users' preference is discussed in Section \ref{sec:personalization}.

\textbf{[C2] to balance \textit{personality} traits:} \citet{thies2017how} observed that users prefer a proactive, productive, witty chatbot. Yet, they also would like them to be caring, encouraging, and exciting. In \citet{mairesse2009can}, the researchers intentionally generated utterances to reflect extreme personalities; as a result, they observed that some utterances sounded unnatural because a human's \textit{personality} is a continuous phenomenon, rather than a discrete one. \citet{sjoden2011extending} also points out that, although \textit{personality} is consistent, moods and states of mind constantly vary. Thus, balancing the predictability of the \textit{personality} and the expected variation is a challenge to overcome.

We also identified strategies to design chatbots that manifest \textit{personality}:

\textbf{[S1] to use appropriate language: }\citet{shum2018eliza} and \citet{morris2002conversational} suggest that the chatbot language should be consistently influenced by its \textit{personality}. Both studies propose that chatbots' architecture should include a persona-based model that encodes the \textit{personality} and influences the response generation. \citet{mairesse2009can} proposed a framework to shows that it is possible to automatically manipulate language features to manifest a particular \textit{personality} based on the Big Five model. The Big Five model is a relevant tool because it can be assessed using validated psychological instruments \citep{mccrae1987validation}. Using this model to represent the \textit{personality} of chatbots was also suggested by \citet{morris2002conversational} and \citet{sjoden2011extending}. \citet{jain2018evaluating} discussed that a chatbot's \textit{personality} should match its domain. Participants expected the language used by the news chatbot to be professional, while they expected the shopping chatbot to be casual and humorous. The ability to use consistent language is discussed in Section \ref{sec:thoroughness}.

\textbf{[S2] to have a sense of humor:} literature highlights humor as a positive \textit{personality} trait \citep{meany2010humour}. In \citet{jain2018evaluating}, ten participants mentioned enjoyment when the chatbots provided humorous and highly diverse responses. The authors found occurrences of the participants asking for jokes and expressing delight when the request was supported. \citet{brandtzaeg2017people} present similar results when arguing that humor is important even for task-oriented chatbots when the user is usually seeking for productivity. For casual conversations, \citet{thies2017how} highlight that timely, relevant, and clever wit is a desired \textit{personality} trait. In \citet{ptaszynski2010towards}, the joker chatbot was perceived as more human-like, knowledgeable, and funny, and participants felt more engaged. 

\textit{Personality} for artificial agents has been studied for some time in the Artificial Intelligence field \citep{elliott1994esearch, rousseau1996personality, petta1997create}. Thus, further investigations on chatbots' \textit{personality} can leverage models for \textit{personality} and evaluate how they contribute to believability and rapport building.

\begin{framed}\small
\vspace{-2mm}
In summary, the \textbf{personification} category includes characteristics that help a chatbot to manifest personal and behavioral traits. The 
benefits relate to increasing believability, human-likeness, engagement, and interpersonal relationship, which is in line with the benefits of \textbf{social intelligence}. However, unlike the \textbf{social intelligence} category, designers and researchers should focus on attributing recognizable \textit{identity} and \textit{personality} traits that are consistent with users' expectations and the chatbot's capabilities. In addition, it is important to care about adaptation to users' cultural context and reducing the effects of negative stereotypes.
\vspace{-2mm}
\end{framed}

\section{Discussion}

In this section, we discuss cross-cutting aspects of the results, namely the influence of the chatbots' perceived humanness, the conversational domains, and the relationships among the characteristics.

\subsection{A note on chatbots' humanness}

The social characteristics identified in the survey align well with characteristics that are present in human-human interactions. On the one hand, this survey shows the benefits of considering these characteristics when developing chatbots, which is supported by the Media Equation theory \citep{nass2000machines, nass1993anthropomorphism}; we identified the domains in which each characteristic was studied. On the other hand, previous studies have also shown that developing overly humanized agents results in high expectations and uncanny feelings \citep{gnewuch2017towards, ciechanowski2018shades}, which was also pointed out in the surveyed literature as a challenge to conveying particular characteristics, such as \textit{identity} and \textit{personality}.

An explanation for these contrasts is that interactivity is ``dependent on the identity of the source with whom/which we are carrying out the message exchange''~\citep{sundar2016theoretical}, i.e., people direct their messages to an artificial agent. In interactions with chatbots, the artificial agent represents the social role usually associated with a human, for example, a tutor~\citep{dyke2013towards, tegos2016investigation}, a healthcare provider~\citep{montenegro2019survey}, a salesperson~\citep{zhu2018lingke, gnewuch2017towards}, a hotel concierge~\citep{lasek2013chatbots}, or even a friend \citep{thies2017how, shum2018eliza}. The idea of assigning a social role to a chatbot does not imply deceiving people into thinking the software is human. Still, as the chatbots enrich their communication and social skills, their perceived social role may approach human profiles.

\subsection{The importance of conversational context}

As Table \ref{tab:domain-analysis} depicts\footnote{see more details in the supplementary materials}, almost all social characteristics (10 out of 11) were found in studies related to conversations not restricted to a specific domain (open domain): manners, moral agency, and emotional intelligence are particularly relevant, since  many conversations involve personal content (e.g., preferences, personal life, and habits); damage control is necessary since interactions are more susceptible to testing or flaming; identity, personality, and thoroughness are adopted to increase believability and consistency; and conscientiousness, proactivity, and personalization foster user engagement and a coherent conversational flow. The only characteristic that was not investigated in an open domain was communicability. As the conversations in this context mimic human social interactions, conveying the chatbots' capabilities is less of a concern.

The reported characteristics vary across task-oriented domains. Education and customer services literature report the greatest number of characteristics, but this result might be influenced by the maturity of the research in chatbots for these domains. As we pointed out in Section \ref{sec:overview}, education and customer services are the task-oriented domains most reported in the literature. We found conscientiousness, damage control, thoroughness, manners, emotional intelligence, and identity in studies for both domains. However, manners and emotional intelligence have a different goal in these domains. In the education context, these characteristics are designed to encourage students, especially in a situation of failure, in which the chatbot should be comforting and sensitive. This function aligns with other domains, such as healthcare. The education domain also reports needs for personality, so the chatbot can be recognized as either an instructor or a student, and proactivity, so the chatbot can motivate students to participate in the interactions. Personality is also reported in other domains in which the chatbots' character influences the interactions, such as gaming and humorous talk. Proactivity is also consistently reported in domains in which the chatbot provides guidance, such as coaching, health, ethnography, and assessment interviews.

In customer services, emotional intelligence and manners are used as a means to manage customers' frustrations with either a product or service. This function is also reported in other domains, such as virtual assistants and information search. Additionally, communicability is important to convey the services provided by the chatbot, and personalization helps to provide services that align with a particular customer, which is observed in domains such as ethnography, task management, virtual assistants, and financial services. Noticeably, personalization mostly co-occurs with proactivity, since users expects proactive messages to be relevant to their particular interactional context.
Moral agency is the only characteristic that does not show up across several domains. It was reported only in race-talk, and open domain. This highlights that moral agency is a concern when the topic involves content that might lead to inappropriate responses. However, as this survey focuses on characteristics reported in the literature, we believe that it might be influenced by the approach adopted by the researchers: it makes sense to study moral agency in the context of gender or race conversations. We argue, though, that this characteristic might be relevant in domains that were not reported here, but that involve interactions that may lead to biased conversations.

\subsection{Interrelationships among the characteristics}
\label{sec:interrelationship}

\begin{figure}[!hb]
  \centering
  \includegraphics[width=8cm]{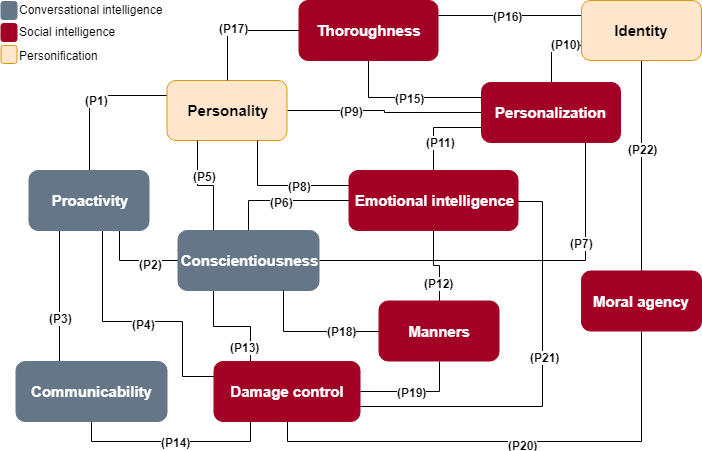}
  \caption{Interrelationship among social characteristics}
  \label{fig:relationship}
\end{figure}

In Section \ref{sec:socialcharacteristics}, we organized the social characteristics into discrete groups and discussed several instances of characteristics influencing each other, or being used as a strategy to manifest one another. In this section, we describe these relations in a theoretical framework, depicted in Figure \ref{fig:relationship}. Rather than providing a comprehensive map, the goal here is to make explicit the relationships found in the surveyed chatbots literature, which underlines the complexity of developing chatbots with appropriate social behaviors. In Figure \ref{fig:relationship}, boxes represent the social characteristics and the colors group them into their respective categories. The axes represent the 22 propositions we derived from the literature.
 
According to the surveyed literature, \textit{proactivity} influences perceived \textit{personality} \textbf{(P1)} \citep{thies2017how}, since recommending and initiating topics may manifest higher levels of extraversion. When the proactive messages are based on the context, \textit{proactivity} increases perceived \textit{conscientiousness} \textbf{(P2)} \citep{thies2017how}, since the proactive messages may demonstrate attention to the topic. \textit{Proactivity} supports \textit{communicability} \textbf{(P3)} \citep{chaves2018single, silvervarg2013iterative, valerio2017here}, since a chatbot can proactively communicate its knowledge and provide guidance; in addition, \textit{proactivity} supports \textit{damage control} \textbf{(P4)} \citep{silvervarg2013iterative}, since a chatbot can introduce new topics when the user either is misunderstood, tries to break the system, or sends an inappropriate message.

\textit{Conscientiousness} is itself a dimension of the Big Five model; hence, \textit{conscientiousness} influences the perceived \textit{personality} \textbf{(P5)} \citep{goldberg1990alternative}. Higher levels of context management, goal-orientation, and understanding increase the chatbots' perceived efficiency, organization, and commitment \citep{dyke2013towards}. \textit{Conscientiousness} manifests\textit{ emotional intelligence} \textbf{(P6)} since retaining information from previous turns and being able to recall them demonstrates empathy \citep{jain2018evaluating}. In addition, \textit{conscientiousness} manifests \textit{personalization} \textbf{(P7)} \citep{jain2018evaluating, thies2017how} because a chatbot can remember individual preferences within and across sessions.

\textit{Emotional intelligence} influences perceived \textit{personality} \textbf{(P8)}, since chatbots' \textit{personality} traits affect the intensity of emotional reactions \citep{morris2002conversational, thies2017how}. Agreeableness is demonstrated through consistent warm reactions such as encouraging and motivating 
\citep{sjoden2011extending, ayedoun2017communication, kumar2010socially, shum2018eliza}. Some \textit{personality} traits 
require \textit{personalization} \textbf{(P9)} to adapt to the interlocutors' culture and interests \citep{ptaszynski2010towards}. Besides, \textit{personalization} benefits \textit{identity} \textbf{(P10)}, since the users' social-agent orientation may require a chatbot to adapt the level of engagement in small talk and the agent's visual representation \citep{jenkins2007analysis, liao2016what}. \textit{Personalization} also improves \textit{emotional intelligence} \textbf{(P11)}, since a chatbot should dynamically regulate the affective reactions to the interlocutor \citep{thies2017how, shum2018eliza}. \textit{Emotional intelligence} improves perceived \textit{manners} \textbf{(P12)}, since the lack of \textit{emotional intelligence} may lead to the perception of impoliteness \citep{jenkins2007analysis}.

\textit{Conscientiousness} facilitates \textit{damage control} \textbf{(P13)}, since the attention to the 
workflow and context may increase the ability to recover from a failure without restarting the workflow \citep{duijst2017can, dyke2013towards, gnewuch2017towards, jain2018convey}. \textit{Communicability} facilitates \textit{damage control} \textbf{(P14)}, since it teaches the user how to communicate, reducing the numbers of mistakes \citep{wallis2005trouble, silvervarg2013iterative, jain2018evaluating, gnewuch2017towards, duijst2017can}. In addition, suggesting how to interact can reduce frustration after failure scenarios \citep{liao2018all}.

\textit{Personalization} manifests \textit{thoroughness} \textbf{(P15)} \citep{thies2017how, gnewuch2017towards, hill2015real, duijst2017can}, since chatbots can adapt their language use to the conversational context and the interlocutor's expectations. When the context requires dynamic variation \citep{thies2017how, gnewuch2017towards}, \textit{thoroughness} may reveal traits of the chatbot's \textit{identity} \textbf{(P16)} \citep{marino2014racial, schlesinger2018let}. As demonstrated by \citep{mairesse2009can}, \textit{thoroughness} also reveals \textit{personality} \textbf{(P17)}.

\textit{Manners} influence \textit{conscientiousness} \textbf{(P18)} \citep{wallis2005trouble}, since they can be applied as a strategy to politely refuse off-topic requests and to keep the conversation on track. \textit{Manners} also influence \textit{damage control} \textbf{(P19)} \citep{curry2018metoo, wallis2005trouble}, because they can help a chatbot prevent verbal abuse and reduce the negative effect of lack of knowledge. Both \textit{moral agency} \textbf{(P20)} and \textit{emotional intelligence} \textbf{(P21)} improve \textit{damage control} because they provide
the ability to appropriately respond to abuse and testing \citep{wallis2005trouble, silvervarg2013iterative}. \textit{Identity} influences \textit{moral agency} \textbf{(P22)}, since \textit{identity} representations require the ability to prevent a chatbot from building or reinforcing negative stereotypes \citep{schlesinger2018let, brahnam2012gender, marino2014racial}.

\section{Related Surveys}
\label{sec:relatedsurveys}

Previous studies have reviewed the literature on chatbots. Several surveys discuss chatbots' urgency \citep{dale2016return, pereira2016chatbots} and their potential application for particular domains, which include education \citep{shawar2007chatbots, deryugina2010chatterbots, rubin2010artificially, satu2015review, winkler2018unleashing}, business \citep{shawar2007chatbots, deryugina2010chatterbots}, health \citep{fadhil2018can, laranjo2018conversational}, information retrieval and e-commerce \citep{shawar2007chatbots}. Other surveys focus on technical design techniques \citep{ramesh2017survey, thorne2017chatbots, winkler2018unleashing, ahmad2018review, walgama2017chatbots, deshpande2017survey, masche2017review}, such as language generation models, knowledge management, and architectural challenges. Although \citep{augello2017overview} discuss the social capabilities of chatbots, the survey focuses on the potential of available open source technologies to support these skills, highlighting technical hurdles rather than social ones.

We found three surveys \citep{pereira2016chatbots, radziwill2017evaluating, ferman2018towards} that include insights about social characteristics of chatbots, although none of them focus on this theme. \citet{pereira2016chatbots} investigated chatbots that ``\textit{mimic conversation rather than understand it},'' and review the main technologies and ideas that support their design, while \citet{ferman2018towards} focuses on identifying best practices for developing script-based chatbots. \citet{radziwill2017evaluating} review the literature on quality issues and attributes for chatbots. The supplementary materials include a table that shows the social characteristics covered by each survey. These related surveys also point out technical characteristics and attributes that are outside the scope of this survey.

\section{Limitations}
\label{sec:limitations}

This research has some limitations. Firstly, since this survey focused on disembodied, text-based chatbots, the literature on embodied and speech-based conversational agents was excluded. We acknowledge that studies that include these attributes can have relevant social characteristics for chatbots, especially for characteristics that could be highly influenced by physical representations, tone, accent, and so forth (e.g. identity, politeness, and thoroughness). However, embodiment and speech could also bring new challenges (e.g., speech-recognition or eye-gazing), which are beyond the scope of this study and could potentially impact users' experiences with chatbots. Additionally, this decision may have caused the under-representation of certain research domains. For example, there is an established research branch on conversational agents in the Ambient Intelligence discipline~\citep{stefanidi2019parlami, leonidis2017using, lopez2010multimodal} that primarily investigates multimodal interactions, and hence is not represented in this survey. Another example is the Software Engineering domain (see, e.g.,~\citep{lebeuf2017software, lin2016developers, abdellatif2020challenges}), where chatbots have been widely investigated, but focused on the functional, rather than social aspects of the interactions.

Secondly, since the definition of chatbot is not consolidated in the literature and chatbots have been studied across several different domains, some studies that include social aspects of chatbots may not have been found. To account for that, we adopted several synonyms in our research string and used Google Scholar as search engine, which provides a fairly comprehensive indexing of the literature in more domains. Finally, the conceptual model of social characteristics was derived through a coding process inspired by qualitative methods, such as Grounded Theory. Like any qualitative coding methods, it relies on the researchers' subjective assessment. To mitigate this threat, the researchers discussed the social characteristics and categories during in-person meetings until reaching consensus, and both the conceptual framework and the relationships amongst the characteristics were derived based on outcomes the surveyed studies explicitly reported.

\section{Conclusion}
\label{sec:conclusion}

In this survey, we investigated the literature on disembodied, text-based chatbots to answer the question \textit{``What chatbot social characteristics benefit human interactions and what are the challenges and strategies associated with them?''} Our main contribution is the conceptual model of social characteristics, from which we can derive conclusions about several research opportunities. Firstly, we point out several challenges to overcome in order to design chatbots that manifest each characteristic. Secondly, further research may focus on assessing the extent to which the identified benefits are perceived by users and influence their satisfaction. Finally, further investigations may propose new strategies to manifest particular characteristics. In this sense, we highlight that we could not identify strategies to manifest moral agency and thoroughness, although strategies for several other characteristics are also under-investigated.

We reported domains where social characteristics were primarily investigated. We showed that some characteristics are largely influenced by the domain (e.g., \textit{moral agency} and \textit{communicability}), while others are more generally applied (e.g., \textit{manners} and \textit{damage control}). We also discussed the relationship among the characteristics through 22 propositions derived from the surveyed literature, which underline the complexity of developing chatbots with appropriate social behaviors. Although the propositions and social characteristics do not intend to be comprehensive, they cover important aspects of human-human communication. Social sciences such as sociolinguistics and communication studies have a good deal to contribute to the design of chatbots. In the description of the characteristics, we pointed out theories that could guide these investigations, such as the cooperative principle \citep{grice1975logic}, social identity \citep{stets2000identity, brown2000social}, personalization \citep{fan2006personalization}, politeness \citep{brown1987politeness}, and mind perception theories \citep{gray2007dimensions}. For example, \textit{conscientiousness}, \textit{thoroughness}, and \textit{proactivity} might be covered by the cooperative principle \citep{grice1975logic}, and social identity theory \citep{stets2000identity, brown2000social} might explain the relevance of \textit{identity}, \textit{personality}, and \textit{moral agency} for conversational technologies. Our results provide important references for helping designers and researchers find opportunities to advance the human-chatbot interactions field.

\section*{Disclosure statement}

No potential conflict of interest was reported by the authors.

\bibliographystyle{apacite}
\bibliography{refs}



\section*{About the Authors}

\textbf{Ana Paula Chaves} is a Ph.D. Candidate at the Northern Arizona University and a Faculty at the Federal University of Technology - Paraná, Campus Campo Mourão, Brazil. She researches social aspects of human-chatbot interactions as well as technologies for tourism. More information at http://anachaves.pro.br

\textbf{Marco Aurelio Gerosa} is an Associate Professor at the Northern Arizona University. He researches Software Engineering and CSCW, having served on the program committee of important conferences, such as FSE, CSCW, SANER, and MSR. He is an ACM and IEEE senior member. For more information, visit http://www.marcoagerosa.com

\end{document}